\documentclass[journal, 12pt, draftclsnofoot, onecolumn]{IEEEtran}
\IEEEoverridecommandlockouts

\usepackage{amsfonts,amssymb}
\usepackage{ifpdf}
\usepackage{cite}
\usepackage{stfloats}
\usepackage{bm}
\usepackage{color,xcolor}
\usepackage{subfigure}
\ifCLASSINFOpdf
   \usepackage[pdftex]{graphicx}
   \graphicspath{{../pdf/}{../jpeg/}}
   \DeclareGraphicsExtensions{.pdf,.jpeg,.png}
\else
   \usepackage[dvips]{graphicx}

   \graphicspath{{../eps/}}

   \DeclareGraphicsExtensions{.eps}
\fi

\usepackage[cmex10]{amsmath}

\usepackage{algorithm}
\usepackage{algorithmic}
\ifCLASSOPTIONcompsoc
\else
  \usepackage[caption=false,font=footnotesize]{subfig}
\fi

\usepackage{makecell}
\begin{document}
\title{ Throughput Maximization for IRS-Aided MIMO FD-WPCN with Non-Linear EH Model}
	
\author{Meng~Hua,
	Qingqing~Wu,~\IEEEmembership{Member,~IEEE}
	%Luxi~Yang,~\IEEEmembership{Senior Member,~IEEE,}
	%Qingqing~Wu,
	%Cunhua~Pan,
	%Chunguo~Li,~\IEEEmembership{Senior Member,~IEEE,}
	%and~A. Lee Swindlehurst,~\IEEEmembership{Fellow,~IEEE}
	%%
	%%%\thanks{Copyright (c) 2015 IEEE. Personal use of this material is permitted. However, permission to use this material for any other purposes must be obtained from the IEEE by sending a request to pubs-permissions@ieee.org.}
	%%\thanks{Manuscript received April   27, 2019; revised July    29, and accepted November  7, 2019. This work was supported by National Natural Science Foundation of China under Grant  61971128,  Grant 61372101, and Grant 61671144, Scientific Research Foundation of Graduate School of Southeast University  under Grand  YBPY1859 and China Scholarship Council (CSC) Scholarship, National High Technology Project of China  under 2015AA01A703,  Cyrus Tang Foundation Endowed Young Scholar Program under SEU-CyrusTang-201801.   The associate editor coordinating the review of this paper and approving it for publication was Kamel Tourki. (\emph{Corresponding author: Luxi Yang}.)}
	\thanks{M. Hua and Q. Wu are with the State Key Laboratory of Internet of Things for Smart City, University of Macau, Macao 999078, China (email: menghua@um.edu.mo; qingqingwu@um.edu.mo). }
}
\maketitle
\vspace{-0.8cm}
\begin{abstract}
This paper studies an intelligent reflecting surface (IRS)-aided  multiple-input-multiple-output (MIMO) full-duplex (FD) wireless-powered communication network (WPCN), where a hybrid access point (HAP) operating in FD broadcasts energy signals to multiple  devices for their energy harvesting (EH)  in the downlink (DL) and meanwhile receives   information  signals  from  devices  in the uplink (UL) with the help of an IRS.  Taking into account the practical finite self-interference (SI) and the non-linear EH model, we formulate  the weighted sum throughput maximization optimization problem   by jointly optimizing     DL/UL time allocation,   precoding matrices at devices,  transmit covariance matrices  at the HAP, and     phase shifts at the IRS. Since the resulting optimization problem is non-convex, there are no standard methods to solve it optimally in general. To tackle this challenge, we first  propose an element-wise (EW) based algorithm, where   each   IRS phase shift is  alternately  optimized  in an iterative manner. To reduce the computational complexity, a   minimum mean-square error (MMSE) based algorithm is proposed, where we transform the original problem  into an equivalent form based on the MMSE method, which   facilities the design of an efficient iterative algorithm. In particular, the IRS phase shift optimization problem is recast as an second-order cone program (SOCP), where all  the IRS phase shifts are simultaneously optimized. For comparison, 
 we also study two suboptimal IRS beamforming configurations in simulations, namely partially dynamic IRS beamforming (PDBF) and static  IRS beamforming (SBF),  which strike a balance between the system performance and practical complexity.  
% Simulation results show that the EW-based algorithm outperforms the MMSE-based algorithm in terms of system throughput,  while the  MMSE-based algorithm achieves  much lower computational complexity from theoretical analysis. 
Simulation results demonstrate the effectiveness of  proposed two  algorithms. Besides, the results  show the superiority of   our proposed scheme 
over other benchmark schemes and also unveil  the importance of the joint design of passive beamforming and resource allocation 
for achieving energy efficient MIMO FD-WPCNs. 
\end{abstract}
\begin{IEEEkeywords}
Intelligent reflecting surface,   full-duplex, WPCN, MIMO, passive  beamforming,  resource allocation.
\end{IEEEkeywords}

\section{Introduction}
Wireless communication via radio-frequency
(RF) radiation has   significantly shaped the people  daily life  in the past decades.  Wireless radio is  not limited   to  forward information-carrying signal  transmission and  its  broadcasting signals    can also be utilized by   Internet-of-Things (IoT) devices, such as temperature sensors, humidity sensors, and  illuminating light sensors, etc.,  to power  the embedded battery \cite{Krikidis2014simultaneous}.  This RF transmission enabled wireless energy transfer (WET) has attracted increasing attention due to its  great convenience as well as perpetual supply compared to the intermittent ambient energy harvesting (EH) technologies such as solar, thermal, vibration, etc. 
 There are mainly two  paradigms of research  on  WET in the literature.  One  is the    simultaneous wireless
information and power transfer (SWIPT),  where   IoT devices can receive information and energy simultaneously  via the same transmitted signal.  The other is the  wireless-powered communication network (WPCN), where  IoT devices  communicate with a hybrid access point (HAP) using the energy harvested from WET. However, for the WPCN, it suffers a ``doubly-near-far'' phenomenon, where  a  device far from the HAP  receives less wireless energy than a nearer user in the downlink (DL), and has to consume more power in the uplink (UL) for wireless information transmission (WIT) \cite{Ju2014Throughput}. Therefore, the low efficiencies of WET and WIT for IoT devices over long transmission
distances fundamentally limit the performance of practical WPCNs.

To address this ``doubly-near-far'' issue, intensive research efforts have been made to 
focus on the design of WPCNs from several  aspects  such as resource allocation, multiple-input-multiple-output (MIMO),  and HAP operation modes in the past. For example,  references \cite{Chingoska2016resource} and  \cite{wqu2018Spectral} studied the joint design of time allocation and transmit power to maximize the sum throughput in WPCNs. 
However, the above works assumed that  the HAP as well as users  is equipped with single antenna, which limits both the DL and UL transmission efficiency. 
%For multiple access schemes, researchers mainly focus on the  UL WIT access. The results in  \cite{Chingoska2016resource} and  \cite{Diamantoulakis2016Wireless} showed that   the non-orthogonal multiple access (NOMA) scheme significantly outperforms the time-division  multiple access (TDMA) scheme in terms of user fairness and/or  sum throughput.
%However, by considering the circuit power consumption at devices,  the results in \cite{wqu2018Spectral} found that the  NOMA-based  scheme  not only consumes more energy, but also is less spectral efficient than the TDMA-based scheme. 
 To exploit   large array gains   brought by multiple antennas, the MIMO technology  is then integrated in WPCNs, where  the HAP and devices are equipped with multiple  antennas. To be specific, with multiple antennas    at the HAP, a narrow beam can be  created and steered  towards to devices  for their EH with  high beamforming gains, and the data transmission rate from devices to the HAP  can be remarkably increased due to the additional spatial  multiplex/diversity gain \cite{Lee2016sum}, \cite{Throughput2015yang}. Although the   MIMO  WPCN    is expected to   achieve much  higher spectral efficiency than the traditional single-input-single-output (SISO)  WPCN, substantial works assume that the operations for   DL WET and  UL WIT are orthogonal over time, i.e, the HAP operates in a half-duplex (HD) mode, which fundamentally  limits the  system performance. To further improve the utilization of the time  resource,   the full-duplex (FD) WPCN was proposed in \cite{ju2014optimal,kang2015full,rezaei2019Secrecy, Chalise2017beamforming}, where the HAP operates in an FD mode.  In particular, reference \cite{ju2014optimal} was the first work to  study SISO FD-WPCN, where  the authors proposed  an efficient transmission  protocol to support simultaneous  DL WET and UL WIT and also studied the weighted sum throughput maximization problem by jointly optimizing  time allocation and power transmit. The  MIMO FD-WPCN was further  studied in \cite{Chalise2017beamforming}, where the authors focused on  a point-to-point system to achieve its  rate-region profile by the joint design of the base station   beamformer and the time splitting.
  However, such above techniques cause huge energy consumption and the green technique for realizing a sustainable wireless network evolution is imperative.

Recently, intelligent reflecting surfaces (IRSs)  have been proposed as a promising cost-effective solution to improve both spectral- and energy-efficiency of  wireless communication systems\cite{wu2020intelligentarxiv,marco2020smart,WU2020towards}. The IRS is  a two-dimensional  planar array comprising a large number of  of sub-wavelength metallic or dielectric scatterers, each of  which is able to independently  manipulate impinging
signals by changing their  phases and/or amplitudes so that the signals
reflected by the IRS can be coherently added up at desired receivers to boost the received power, while  destructively added up at non-intended receivers to suppress co-channel interference. In  particular,  massive reflecting elements can be installed at the IRS due to the small size  of each reflecting element, which are  able to provide significant passive gains to compensate  the high signal attenuation over long distance \cite{WU2020towards}. The seminal work in \cite{wu2019intelligentxx} unveiled the fundamental scaling law of the  IRS by showing that the received signal-to-noise ratio (SNR) increases quadratically with the number of IRS reflecting elements. Inspired by this fundamental insight, IRS-aided wireless communications have been intensively studied for  various system setups  including physical layer security \cite{Intelligent2020guan,zhang2020robust,yu2020robust}, multi-cell cooperation \cite{pan2020multicell,hua2020intelligent,xie2020max},  non-orthogonal multiple access (NOMA) \cite{zhu2021power,ding2020simple,chen2021irsaided}.
While most of existing works focus on the  IRS-aided WIT, the  IRS is  also appealing for WET \cite{wu2019weighted,pan2019intelligent,li2020joint,zargari2021max}.  First, 
due to the large aperture of IRS,  the signals reflected by  the properly reconfigured  IRS can be coherently added up with the non-IRS reflected signals  at the desired devices/HAP to substantially  boost     DL WET and UL WIT. Second, the energy consumption and the hardware cost of IRS are very low and cheap since  the IRS  reflecting elements   only passively reflect  impinging signals without requiring any transmit RF chains. Thus, its energy consumption is typically  several  orders-of-magnitude lower than that of  traditional active antenna arrays  \cite{tang2021wireless}.
% Compared to the MIMO technology,  the  IRS not only    provides considerable system performance gain, but  also provides a more greener solution to the future network.

By far,  there are    only a handful of works paying attention to studying IRS-aided  WPCNs
\cite{wu2021irsxx,9610992,wu2021irs,lyu2021OptimizedEnergy,zhang2021throughput,gong2021Throughput,chu2021novel,zheng2020Intelligent}. For example, a new dynamic IRS beamforming for WPCNs was proposed to compromise the system  performance and implementation complexity in   \cite{wu2021irsxx}. 
The results in these works showed that with the joint optimization of IRS phase shifts and resource allocation, the system performance can be significantly improved.  However,  these works adopted the \textit{linear} energy harvesting (EH) model for DL WET and  ignored the EH saturation caused  by  practically \textit{non-linear} rectifiers \cite{Bruno2019fundamental}.  Specifically, it was shown that    the output direct current (DC) power stored in the embedded battery at   devices does not linearly scale  with  the input RF power, especially   as  the input RF power level  increases,  the output DC power will become saturated gradually. As a result, employing a linear EH model may lead to mismatched beamforming and resource allocation results for WPCNs with practical non-linear rectifiers.
Moreover,  all  above works  focused on the  HD system, which suffers from    low spectral  efficiency. We note that although the FD technique   applied in IRS-aided wireless communication networks has been  studied in \cite{xu2020resource,shen2020beamforming,cai2020intelligent}, these works 
 assumed that  the self-interference (SI) at the HAP is  either a constant or is  perfectly  cancelled    by ignoring the practical  quantization
 error introduced by the strong SI under the limited dynamic range of the analog-to-digital
 conversion. The  research on the joint design of  FD and IRS technologies under practical constraints in MIMO WPCNs are   still an open problem, which thus
 motivates this work.
 
 \begin{figure}[!t]
 	\centerline{\includegraphics[width=3in]{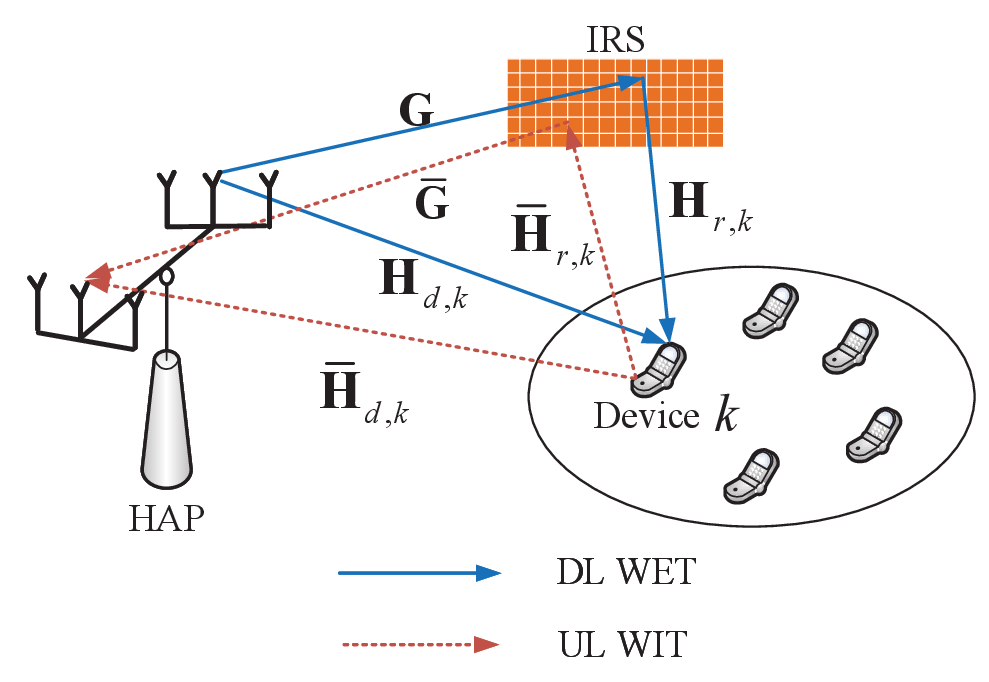}}
 	\caption{An IRS-aided MIMO FD-WPCN.} \label{fig1}
 	\vspace{-0.5cm}
 \end{figure} 
In this paper, we study an IRS-aided MIMO FD-WPCN  for further improving the system throughput as shown in Fig.~\ref{fig1}. The HAP operates in the FD mode and is equipped with two sets of antennas, which are used for DL WET and UL WIT, respectively, and  each  device is equipped with multiple antennas. The devices harvest energy from the HAP and  use the harvested energy to  transmit their own information to the HAP  with the help of IRS. 
% In addition, the IRS is deployed close  to  the devices. 
%With  configurations of multiple antennas as well as  the  IRS, the spatial multiplexing  and active/passive beamforming  gains can be exploited to enhance the system performance.  
To the best of our knowledge, it is the first work to study  the IRS-aided  MIMO FD-WPCN by taking  into    account the  finite SI and the non-linear EH model. 
Our work is significantly different from the  IRS-aided SISO FD-WPCN proposed in \cite{hua2021joint} mainly from three aspects. First, the extension from the SISO system to the MIMO system resulting in a different problem formulation, thus we propose two new  algorithms to cater to the new formulated problem.
Second, we adopt a more practical EH model, i.e., the non-linear EH model, rather than the ideal EH model, i.e., the linear EH model, to capture the dynamics of the RF energy conversion efficiency for different input power levels. Third, we propose  a novel  DL transmission protocol, where  a  dynamic IRS beamforming (DBF) strategy is proposed to adapt to  different  channel fading among devices  in the  DL WET. Moreover,  compared to the IRS-aided  MIMO HD-WPCN \cite{gong2021Throughput},  our work      differs significantly in terms of   transmission protocols,  transmitter/receiver architectures, EH  constraints, design objectives, and optimization methods. 
%It is worth pointing out that our work is significantly different from the previous work \cite{gong2021Throughput} from five  aspects. 
%First, work \cite{gong2021Throughput} considers an IRS-aided MIMO WPCN with the HAP operating in an HD mode, while our work consider that the HAP operates in an FD mode considering finite SI.  
% Second,  work \cite{gong2021Throughput} considers that   one common IRS phase-shift vector is designed for the whole  DL WET stage, while a fully dynamic IRS beamforming is assumed to adapt to time-selective channel fading. 
%% Specifically, the time duration of DL WET is further divided into multiple time sub-slots, and  multiple IRS phase-shift vectors are designed and vary with each time sub-slot for DL WET to adapt to different channels among  different devices. 
%Third, different from  work \cite{gong2021Throughput} considering the ideal EH model, i.e., linear EH model, a practical EH model, i.e., the non-linear EH model,  is adopted to capture the dynamics of
%the RF energy conversion efficiency for different input power levels.
%Fourth, to balance the system performance and implementation complexity, the  static IRS beamforming is also analyzed and compared in this paper. Fifth, due to the different problem formulations, the proposed algorithm in \cite{gong2021Throughput} cannot be applied to our problem formulation, we propose two new  algorithms to cater to the new formulated problem.
The main contributions of this paper are summarized as follows.
\begin{itemize}
	\item We study  an IRS-aided MIMO FD-WPCN considering   practical cases with the  finite SI and the non-linear EH model. Our objective  is to maximize the weighted sum throughput by jointly optimizing  DL/UL time allocation,   precoding matrices at devices, transmit covariance matrices at  the HAP, and phase shifts at the IRS.   Since the resulting optimization problem is non-convex, there are no standard convex methods to solve it optimally. To efficiently solve it, two different algorithms, namely    element-wise (EW) based algorithm and   minimum mean-square error (MMSE) based algorithm, are proposed to strike a balance between the system performance and the computational complexity.
	
	\item For the proposed  EW-based algorithm, the original optimization problem is decomposed into several subproblems, namely  DL/UL time allocation  subproblem,   precoding matrices and transmit covariance matrices  subproblem, and IRS phase shift  subproblem. Then, we alternately optimize each subproblem in an iterative way until   convergence is achieved. In particular, for the IRS phase shift optimization, we alternately optimize each phase shift with other phase shifts are fixed. Although the proposed EW-based algorithm provides a good system performance as shown in simulation results, it requires procedures for  updating each phase shift in a one-by-one manner. 
	
	\item To reduce the computational complexity,  the  MMSE-based algorithm is proposed. Specifically, the original optimization  problem is first  equivalently transformed into a more tractable form based on the MMSE method. Based on the newly formulated optimization problem, the IRS phase shift optimization problem is recast as  a standard second-order cone program (SOCP)  by applying the successive convex approximation (SCA) technique, which can be solved with much  lower complexity. 
	
	\item Simulation results  demonstrate the benefits of the    IRS  used for enhancing the performance of the MIMO FD-WPCN, especially when the  DBF scheme is adopted for both DL WET and UL WIT. In addition, the   IRS-aided MIMO FD-WPCN is able to achieve significantly  performance gain compared to the  IRS-aided MIMO HD-WPCN when the SI is well suppressed. Furthermore, it is interesting to find that while the DBF is useful for UL WIT,  it may not be necessary for DL WET, which helps reduce the implementation complexity   in practice.	
\end{itemize}
The rest of this paper is organized as follows. Section II introduces the system model and problem formulation  for the IRS-aided MIMO FD-WPCN. In Section III, we propose an EW-based algorithm. In Section IV, we propose an MMSE-based algorithm. 
 Numerical results are provided in Section V and  the paper is concluded in Section VI.

\emph{Notations}: Boldface upper-case and lower-case  letters denote matrix and   vector, respectively.  ${\mathbb C}^ {d_1\times d_2}$ stands for the set of  complex $d_1\times d_2$  matrices. For a complex-valued vector $\bf x$,  ${\rm diag}(\bf x) $ denotes a diagonal matrix whose main diagonal elements are extracted from vector $\bf x$.
   For a   matrix $\bf X$, ${\bf X}^*$,  ${\bf X}^H$, and ${\rm{tr}}\left( {\bf{X}} \right)$ respectively  stand for  its  conjugate,  conjugate transpose, and trace, while ${\bf{X}} \succeq {\bf{0}}$ indicates that matrix $\bf X$ is positive semi-definite.
${\bf{X}} \left(\bf x \right) \sim {\cal CN}\left( {{\bm \Upsilon} \left(\bm \mu \right) ,{\bf{\Sigma }}} \right)$ denotes a circularly symmetric complex Gaussian matrix (vector) with mean $\bm \Upsilon \left( \bm \mu \right)$ and covariance matrix $\bm \Sigma$.  ${\mathbb E}(\cdot)$ denotes the expectation operation. $ \odot $ and $ \otimes $ represent the  Hadamard product and Kronecker product, respectively. ${\cal O}\left(  \cdot  \right)$ denotes  the big-O computational complexity notation.

\section{System Model and Problem Formulation}
\subsection{System Model}
Consider an IRS-aided FD-WPCN consisting of an  HAP, $K$  devices, and an IRS, as shown in Fig.~\ref{fig1}.  
We assume that the HAP operates in the FD mode and the total number of antennas at the  HAP is $N_t+N_r$, of which $N_t$ transmit antennas are used
for  broadcasting  energy to the distributed devices in the DL and $N_r$ receive
antennas are dedicated to receiving  information from  the distributed devices in the UL  simultaneously over the same frequency band. In addition, we assume that the 
 IRS has $M$ reflecting elements and  each device has $N_d$ antennas.

 \begin{figure}[!t]
	\centerline{\includegraphics[width=6in]{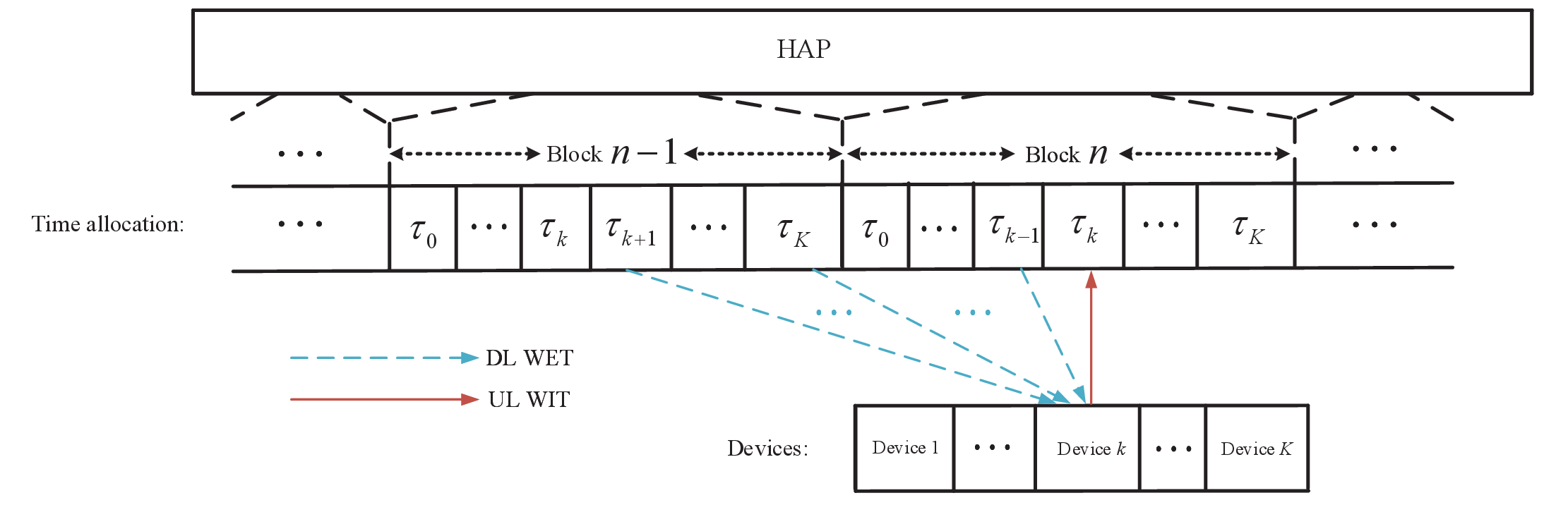}}
	\caption{Transmission protocol  for  DL WET and UL WIT in the considered MIMO FD-WPCN.}\vspace{-0.5cm} \label{fig2}
\end{figure}

\begin{figure}[!t]
	\centerline{\includegraphics[width=3in]{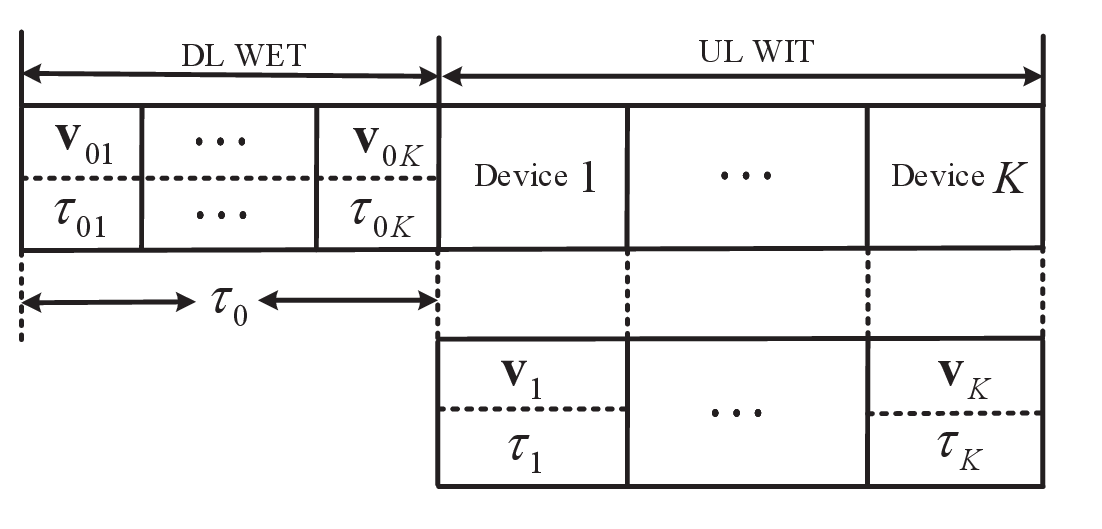}}
	\caption{The DBF scheme for  DL/UL time allocation and IRS phase-shift vectors assignment.} \label{fig3}
	\vspace{-0.5cm}
\end{figure}

We consider a quasi-static flat-fading channel  in which the channel state information (CSI) remains constant in a channel coherence  frame, but may change in the subsequent frames.  As shown in Fig.~\ref{fig2}, each channel coherence frame consists of multiple blocks and each transmission period
of one block of interest denoted by
 $T$  is divided into $K+1$ time  slots, each with duration of $\tau_i,i\in{\cal K}_1$,  such that $T = \sum\limits_{i = 0}^K {{\tau _i}} $,  where ${\cal K}_1=\{0,\dots,K\}$. The $0$th time slot is a dedicated time slot used to broadcast  energy to all distributed devices in the DL, while the  $i$th ($i \ne 0$) time slot is used for UL WIT.  The DBF scheme for DL WET is proposed as shown in  Fig.~\ref{fig3}. Without loss of generality,   the $0$th time slot is further divided into $K$ sub-slots with  the duration of  the $j$th sub-slot denoted by $\tau_{0j},j\in{\cal K}_2$, such that ${\tau _0} = \sum\limits_{j = 1}^{{K}} {{\tau _{0j}}} $,  where ${\cal K}_2=\{1,\dots,K\}$. Then, each sub-slot is  assigned an exclusive IRS phase-shift vector.  For ease of exposition,    we assume that the $k$th $(k\ne0)$ time slot is occupied by device  $ k$ for UL WIT, $k\in{\cal K}_2$.
  Since there are two sets of antennas at the HAP, the DL   and UL channels are   different in general. In the DL transmission, denote by  ${\bf{G}} \in {{\mathbb C}^{M \times N_t}}$, ${\bf{H}}_{r,k} \in {{\mathbb C}^{N_d \times M}}$, and ${{\bf H}_{d,k}} \in {{\mathbb C}^{N_d \times N_t}}$ the complex equivalent baseband   channel between the HAP  and  the  IRS, between  the IRS  and the $k$th device, and  between  the  HAP and the $k$th device,  $k\in{\cal K}_2$, respectively.  In the UL transmission, denote by  ${\bf{\bar G}} \in {{\mathbb C}^{N_r \times M}}$, ${\bf{\bar H}}_{r,k} \in {{\mathbb C}^{M \times N_d}}$, and ${{\bar {\bf H}}_{d,k}} \in {{\mathbb C}^{N_r \times N_d}}$ the complex equivalent baseband   channel between the HAP  and  the IRS, between the IRS  and the $k$th device, and  between  the  HAP and the $k$th device,  $k\in{\cal K}_2$, respectively. In addition, the channel reciprocity is assumed and thus we have   ${{{\bf{\bar H}}}_{r,k}} = {\bf{H}}_{r,k}^T$ for the IRS-device link, $k\in{\cal K}_2$.

\subsubsection{DL WET}
As shown in Fig.~\ref{fig3},  each sub-slot in the $0$th time slot is assigned an exclusive IRS phase-shift vector.  As such,
the received signal  by device $k$ in  the  $i$th sub-slot  of time slot $0$ is given by
\begin{align}
{\bf{y}}_{k,i}^d = \left( {{{\bf{H}}_{d,k}} + {{\bf{H}}_{r,k}}{{\bf{\Theta }}_i^0}{\bf{G}}} \right){{\bf{x}}_i^0} + {\bf{n}}_{k,i}^d, i \in {\cal K}_2, k \in {\cal K}_2,
\end{align}
where ${{\bf{x}}_i^0}$  
represents for the HAP's transmit energy signal in  the  $i$th sub-slot  of time slot $0$  with  covariance matrix  ${{\bf{Q}}_i^0} = {\mathbb E}\left\{ {{{\bf{x}}_i^0}{\bf{x}}_i^{0,H}} \right\}\succeq {\bf{0}} $.\footnote{ Note that ${{\bf{x}}_i^0}$ can be treated as the sum of multiple independent energy beams transmitted by the HAP \cite{xujie2014energy}, i.e., ${\bf{x}}_i^0{\rm{ = }}\sum\limits_{j = 1}^{{{\tilde N}_t}} {{\bf{w}}_{i,j}^0x_{i,j}^0} $,  where ${{{\tilde N}_t}}$ denotes the number of energy beams satisfying ${{{\tilde N}_t}}\le {{N_t}}$, ${{\bf{w}}_{i,j}^0} \in {\mathbb C}^{N_t \times 1}$ denotes the $j$th energy beam and   ${x_{i,j}^0}$ represents  its carried energy-modulated
	signal satisfying  ${\mathbb E}\left( {{{\left| {x_{i,j}^0} \right|}^2}} \right) = 1$. As such, we have ${\bf{Q}}_i^0 = {\mathbb E}\left( {{\bf{x}}_i^0{\bf{x}}_i^{0,H}} \right) = \sum\limits_{j = 1}^{{{\tilde N}_t}} {{\bf{w}}_{i,j}^0{\bf{w}}_{i,j}^{0,H}}$. Obviously, given ${\bf{Q}}_i^0$, the above energy beams can be recovered  from the
	eigenvalue decomposition (EVD) of ${\bf{Q}}_i^0$ with ${{\tilde N}_t} = {\rm{rank}}\left( {{\bf{Q}}_i^0} \right)$. Based on this, we directly optimize ${{\bf{Q}}_i^0}$  in the sequel of this paper.}      ${{\bf{\Theta }}_{i}^0} = {\rm{diag}}\left( {{v_{i,1}^0}, \cdots ,{v_{i,M}^0}} \right)$  represents  a diagonal IRS reflection coefficient matrix  in the  $i$th sub-slot  of time slot $0$, where  ${{v_{i,m}^0}}$ denotes its
 $m$th phase shift, and ${\bf n}_{{k},i}^d \sim {\cal CN} \left( {{\bf 0},\sigma ^2{\bf I}_{N_r}} \right)$ stands for the additive white Gaussian noise. 
  It is worth noting that  we consider a {\it periodic} transmission protocol  as shown in Fig.~\ref{fig2},  where each device $k$  can harvest energy during all time slots except its own transmission time slot $k$\cite{ju2014optimal}.  Specifically, the energy harvested by device $k$   after its own WIT in the previous   transmission block will be used for its WIT in the next transmission block. 
 Under this protocol,  the amount of  harvested energy   by device $k$ during $T$ by using the practical EH model is given by \cite{Boshkovska2015Practical}, \cite{lu2019global}\footnote{ 
 	The energy harvested from the noise and the received UL WIT signals from other
 	devices are assumed to be negligible, since both the noise power and device transmit power are
 	much smaller than the HAP transmit power  in practice \cite{Ju2014Throughput},\cite{wqu2018Spectral}.}  
\begin{align}
{E_k} = &\sum\limits_{i = 1}^K {{\tau _{0i}}\left( {\frac{{{X_k}}}{{1 + {e^{ - {a_k}\left( {{\rm{tr}}\left( {{\bf{H}}_{k,i}^0{\bf{Q}}_i^0{\bf{H}}_{k,i}^{0,H}} \right) - {b_k}} \right)}}}} - {Y_k}} \right)}  +  \notag\\
&\sum\limits_{j = 1,j \ne k}^K {{\tau _j}\left( {\frac{{{X_k}}}{{1 + {e^{ - {a_k}\left( {{\rm{tr}}\left( {{{\bf{H}}_{k,j}}{{\bf{Q}}_j}{\bf{H}}_{k,j}^H} \right) - {b_k}} \right)}}}} - {Y_k}} \right)} ,  k\in{\cal K}_2, 
\end{align}
where ${\bf{H}}_{k,i}^0 = {{\bf{H}}_{d,k}} + {{\bf{H}}_{r,k}}{\bf{\Theta }}_i^0{\bf{G}}$ and ${{\bf{H}}_{k,j}} = {{\bf{H}}_{d,k}} + {{\bf{H}}_{r,k}}{{\bf{\Theta }}_j}{\bf{G}}$. ${\bf Q}_j\succeq {\bf 0}$  stands for  the transmit covariance matrix at the HAP in time slot $j$,  and ${{\bf{\Theta }}_j} = {\rm{diag}}\left( {{v_{j,1}}, \ldots ,{v_{j,M}}} \right)$ denotes the IRS phase-shift matrix assigned in    time slot $j$, $j\in {\cal K}_2$. In addition, 
$a_k$,  $b_k $, $M_k$,   ${X_k} = {M_k}\left( {1 + {e^{{a_k}{b_k}}}} \right)/{e^{{a_k}{b_k}}}$,  and  ${Y_k} = {{{M_k}}}/{{{e^{{a_k}{b_k}}}}}$ represent device $k$'s circuit  configurations, which are 
related to  the resistance, capacitance,  diode turn-on voltage, etc.

\subsubsection{UL WIT}
During UL WIT, each device  transmits its own  to the HAP in a time-division multiple access manner. As such,  the  received signal  at the HAP in time slot $k$ is given by  
\begin{align}
{\bf{y}}_k^r = \underbrace {\left( {{{{\bf{\bar H}}}_{d,k}} + {\bf{\bar G}}{{\bf{\Theta }}_k}{{{\bf{\bar H}}}_{r,k}}} \right){{\bf{W}}_k}{{\bf{s}}_k}}_{{\rm{desired}}\;{\rm{signal}}} + \underbrace {\left( {{{\bf{H}}_{{\rm{SI}}}} + {\bf{\bar G}}{{\bf{\Theta }}_k}{\bf{G}}} \right){{\bf{x}}_k}}_{{\rm{interference}}} + {\bf{n}}_k^r, k\in{\cal K}_2,\label{up_receiver}
\end{align}
where  ${{\bf{x}}_k} \in {\mathbb C}^{N_t \times 1} $  denotes the   energy signal transmitted by the HAP with the transmit covariance matrix denoted by ${{\bf{Q}}_{k}} \succeq {\bf{0}}$ in time slot $k$. 
${{\bf{s}}_k} \in {{\mathbb C}^{{N_d} \times 1}}$ represents $N_d$ desired data streams for device $k$ satisfying ${{\bf{s}}_k} \sim {\cal CN}\left( {{\bf{0}},{{\bf{I}}_d}} \right)$ and  ${{{\bf{W}}_k}} \in {\mathbb C}^{N_d \times N_d}$ stands for the transmit beamforming matrix for device $k$.
 ${{{\bf{H}}_{{\rm SI}}}}\in {\mathbb C}^{N_r \times N_t}$ represents the SI channel from the transmit antennas to the
 receive antennas at the FD-HAP, and  ${\bf n}_{k}^r \sim {\cal CN} \left( {{\bf 0},\sigma ^2{\bf I}_{N_r}} \right)$ denotes the received additive white Gaussian noise.
It can be observed that the interference  in \eqref{up_receiver}  consists of two parts. The first  term ${{\bf{H}}_{\rm SI}}{{\bf{x}}_k}$ denotes the SI resulting from the DL transmission by the HAP  and the second  term ${\bf{\bar G}}{{\bf{\Theta }}_k}{\bf{G}}{{\bf{x}}_k}$ denotes the interference  introduced by the reflection of the DL transmit signal by the IRS. By  deploying the IRS  close to   devices and far from the HAP, the following two benefits can be achieved. First, when the  IRS is deployed close to   devices,  the double propagation loss of  the  cascaded  HAP-IRS-device  link will be substantially reduced, which is beneficial for improving the SNR. Second, when the IRS is deployed far from the HAP, the   interference  introduced by the reflection  link, i.e., HAP-IRS-HAP link,  will be significantly small and thus can be ignored. 
 Thus, we  can simplify \eqref{up_receiver} as  
\begin{align}
{\bf{\hat y}}_k^r = {{{\bf{\bar H}}}_k}{\bf W}_k{{\bf{s}}_k} + {{\bf{H}}_{{\rm{SI}}}}{{\bf{x}}_k} + {\bf{n}}_k^r,k\in{\cal K}_2, \label{up_receiver_new}
\end{align} 
where ${{{\bf{\bar H}}}_k} = {{{\bf{\bar H}}}_{d,k}} + {\bf{\bar G}}{{\bf{\Theta }}_k}{{{\bf{\bar H}}}_{r,k}}$. As a result, the achievable throughput of device $k$   at time slot $k$ can be  expressed as
\begin{align}
{R_k} = {\tau _k}\ln \left| {{{\bf{I}}_{{N_r}}} + {{{\bf{\bar H}}}_k}{{\bf{W}}_k}{{\bf{W}}_k^H}{\bf{\bar H}}_k^H{{\left( {{{\bf{H}}_{{\rm{SI}}}}{{\bf{Q}}_k}{\bf{H}}_{{\rm{SI}}}^H + {\sigma ^2}{{\bf{I}}_{{N_r}}}} \right)}^{ - 1}}} \right|.\label{rate}
\end{align}
\subsection{Problem Formulation}
Define ${{\bf{v}}_{i}^0} = {\left[ {{v_{i,1}^0}, \ldots ,{v_{i,M}^0}} \right]^T}$ and  ${{\bf{v}}_i} = {\left[ {{v_{i,1}}, \ldots ,{v_{i,M}}} \right]^T}, i\in {\cal K}_2$.
 Denote by sets ${\cal M} = \left\{ {1, \ldots ,M} \right\}$, $\tau {\rm{ = }}\left\{ {\tau_{0,k}}\cup  {{\tau _k},k \in {{\cal K}_2}} \right\}$,  ${\bf{W}} = \left\{ {{{\bf{W}}_k},k \in {{\cal K}_2}} \right\}$, ${\bf{v}} = \left\{ {{{\bf{v}}_{i}^0} \cup {{\bf{v}}_i}, i \in {{\cal K}_2}} \right\}$,  and ${\bf{Q}} = \left\{ { {{\bf{Q}}_i^0} \cup  {{\bf{Q}}_i},i \in {{\cal K}_2}} \right\}$. 
 The  objective of this paper is to   maximize the weighted sum  throughput of the IRS-aided MIMO FD-WPCN by jointly optimizing  DL/UL time allocation,    precoding matrices at devices, transmit covariance matrices  at the HAP, and phase-shift vectors at the IRS.    
 Accordingly, the problem is formulated as
% \footnote{The considered sum throughput maximization problem   can be readily extended to the weighted sum throughput maximization problem by   adding   different weighting factors on each device in the objective function, which does not   affect our proposed algorithms.  Without loss of generality,  we focus on the   sum throughput maximization problem.}  
\begin{subequations} \label{p1}
	\begin{align}
	&\mathop {\max }\limits_{{\bf{W}},{\bf{Q}}\succeq{\bf 0},{\bf{v }},\tau \ge 0 } \sum\limits_{k = 1}^K {{\tau _k}{\omega _k}\ln \left| {{{\bf{I}}_{{N_r}}} +  {{{{\bf{\bar H}}}_k}{{\bf{W}}_k}{{\bf{W}}_k^H}{\bf{\bar H}}_k^H} {{\left( {{{\bf{H}}_{{\rm{SI}}}}{{\bf{Q}}_k}{\bf{H}}_{{\rm{SI}}}^H + {\sigma ^2}{{\bf{I}}_{{N_r}}}} \right)}^{ - 1}}} \right|}  \label{p1_obj}\\
	&{\rm s.t.}~{\tau _k}{\rm{tr}}\left( {{\bf{W}}_k}{{\bf{W}}_k^H} \right) \le \sum\limits_{i = 1}^K {{\tau _{0i}}\left( {\frac{{{X_k}}}{{1 + {e^{ - {a_k}\left( {{\rm{tr}}\left( {{\bf{H}}_{k,i}^0{\bf{Q}}_i^0{\bf{H}}_{k,i}^{0,H}} \right) - {b_k}} \right)}}}} - {Y_k}} \right)}  +  \notag\\
	&\qquad \sum\limits_{j = 1,j \ne k}^K {{\tau _j}\left( {\frac{{{X_k}}}{{1 + {e^{ - {a_k}\left( {{\rm{tr}}\left( {{{\bf{H}}_{k,j}}{{\bf{Q}}_j}{\bf{H}}_{k,j}^H} \right) - {b_k}} \right)}}}} - {Y_k}} \right)} , k \in {\cal K}_2,\label{p1_const_1}\\
		& \qquad \sum\limits_{j = 1}^K {{\tau _{0j}}}  + \sum\limits_{k= 1}^K {{\tau _k}}  \le T,\label{p1_const_2}\\
%	&\qquad{\tau _k} \ge 0,{k \in {{\cal K}_1}},\\
	&\qquad {\rm{tr}}\left( {{{\bf{Q}}_k^0}} \right) \le {P_{\max }}, {\rm{tr}}\left( {{{\bf{Q}}_k}} \right) \le {P_{\max }},k \in {{\cal K}_2},\label{p1_const_3}\\
	&\qquad \left| {{{ {{{{v}}_{i,m}^0}} }}} \right| = 1,\left| {{{ {{{{v}}_{i,m}}} }}} \right| = 1, m \in {\cal M}, i   \in {{\cal K}_2}\label{p1_const_4},
	\end{align}
\end{subequations}
where ${\omega _k}$ $\left( {{\omega _k} > 0}, k\in {\cal K}_2 \right)$ denotes the weighting factor for device $k$ and  $P_{\max}$ in \eqref{p1_const_3} denotes the  maximum allowed   transmit power at the HAP. Constraint \eqref{p1_const_1} represents the  energy causality constraint, where the transmitted energy consumption by the device cannot exceed its harvested energy. Constraints \eqref{p1_const_2} and \eqref{p1_const_4} denote the    constraints on the total
transmission time and IRS phase shifts, respectively.
Problem \eqref{p1} is challenging to solve due to the following reasons. First, the optimization variables $\bf W$ and $\bf Q$ are  coupled in the logarithmic form in    \eqref{p1_obj}, which makes the objective function  non-convex. Second,  the optimization variables $\bf v$ and $\bf Q$ are coupled in the  sigmoid function in \eqref{p1_const_1}, which further complicates the problem. Third,  constraint \eqref{p1_const_4} is a unit-modulus constraint, which is also non-convex. In general, there are no standard convex techniques to solve it optimally. To obtain a high-quality solution, we propose two algorithms, namely EW-based algorithm and MMSE-based algorithm, which are  elaborated  in Section~\ref{sectionIII} and Section~\ref{sectionIV}, respectively.
\section {EW-based  Algorithm} \label{sectionIII} 
In this section, we propose an EW-based algorithm. Specifically,  we decompose   problem  \eqref{p1} into several subproblems, and then alternately optimize each subproblem  until   convergence is achieved. In particular, for the IRS phase shift optimization, the phase shifts   are  solved by a one-by-one manner. Define ${{\bf{S}}_k} = {{\bf{W}}_k}{\bf{W}}_k^H \succeq {\bf 0}$ and  ${\bf{S}} = \left\{ {{{\bf{S}}_k}, k\in {\cal K}_2} \right\}$. 
 By introducing  auxiliary variables   $z = \left\{ {{z_{k,i}} \cup z_{k,i}^0},k \in {{\cal K}_2}, i   \in {{\cal K}_2} \right\}$  and  replacing  ${{\bf{W}}_k}{\bf{W}}_k^H$ with ${{\bf{S}}_k}$, problem \eqref{p1} can be equivalently transformed as 
\begin{subequations} \label{p1_1}
	\begin{align}
	&\mathop {\max }\limits_{{\bf{S}}\succeq{\bf 0},{\bf{Q}}\succeq{\bf 0},{\bf{v }},\tau \ge 0 ,z\ge 0} \sum\limits_{k = 1}^K {{\tau _k}{\omega _k}\ln \left| {{{\bf{I}}_{{N_r}}} +  {{{{\bf{\bar H}}}_k}{{\bf{S}}_k}{\bf{\bar H}}_k^H} {{\left( {{{\bf{H}}_{{\rm{SI}}}}{{\bf{Q}}_k}{\bf{H}}_{{\rm{SI}}}^H + {\sigma ^2}{{\bf{I}}_{{N_r}}}} \right)}^{ - 1}}} \right|} \label{p1_1_obj}  \\
	&{\rm s.t.}~\tau_k{\rm{tr}}\left( {{{\bf{ S}}_k}} \right) \le \sum\limits_{i = 1}^K {{\tau _{0i}}\left( {\frac{{{X_k}}}{{1 + z_{k,i}^0}} - {Y_k}} \right)}  + \sum\limits_{j = 1,j \ne k}^K {{\tau _j}\left( {\frac{{{X_k}}}{{1 + {z_{k,j}}}} - {Y_k}} \right)} , k \in {\cal K}_2,\label{p1_1_const_1}\\
	&\qquad z_{k,i}^0 \ge {e^{ - {a_k}\left( {{\rm{tr}}\left( {{\bf{H}}_{k,i}^0{\bf{Q}}_i^0{\bf{H}}_{k,i}^{0,H}} \right) - {b_k}} \right)}},k \in {{\cal K}_2},i \in {{\cal K}_2},\label{p1_1_const_2}\\
	& \qquad {z_{k,j}} \ge {e^{ - {a_k}\left( {{\rm{tr}}\left( {{{\bf{H}}_{k,j}}{{\bf{Q}}_j}{\bf{H}}_{k,j}^H} \right) - {b_k}} \right)}},k \in {{\cal K}_2},j\in {{\cal K}_2},\label{p1_1_const_3}\\
	& \qquad \eqref{p1_const_2},\eqref{p1_const_3},\eqref{p1_const_4}.
	\end{align}
\end{subequations}
It can be readily  verified that  at the optimal solution to problem \eqref{p1_1}, constraints   \eqref{p1_1_const_2} and \eqref{p1_1_const_3} are met with  equalities,
since otherwise one can always
decrease $z_{k,i}^0$ or $z_{k,j}$ to obtain a strictly larger UL WIT duration  $\tau_k$ in \eqref{p1_1_const_1}, which results in a strictly larger objective value. In addition, with the given ${\bf S}_k$, ${{\bf{W}}_k}$ can be obtained from performing  EVD.
Thus, problem \eqref{p1_1} is equivalent to problem \eqref{p1}. 
%With such a reformulation, the introduced new constraints \eqref{p1_1_const_2} and \eqref{p1_1_const_3} are convex, but the new constraint \eqref{p1_1_const_1} is still non-convex. 
To efficiently solve  problem \eqref{p1_1},    we partition  all  variables into three  blocks: 1) precoding matrices and transmit covariance matrices 
$\{{\bf{ S}}, {\bf Q}, z\}$, 2) DL/UL time allocation  $\tau$, and  3) IRS phase shifts $\bf{v }$. Then, we  alternately solve each  block with the other two blocks are fixed, until convergence is achieved.
\subsection{Precoding Matrix and Transmit Covariance Matrix Optimization}
For any given  time allocation  $\tau$  and IRS phase shifts  $\bf{v }$,  the precoding matrices and transmit covariance matrices in  problem \eqref{p1_1} can be optimized by solving the following problem
\begin{subequations} \label{p1_2}
	\begin{align}
	&\mathop {\max }\limits_{{\bf{S}}\succeq{\bf 0},{\bf{Q}}\succeq{\bf 0},z\ge 0} \sum\limits_{k = 1}^K {{\tau _k}{\omega _k}\ln \left| {{{\bf{I}}_{{N_r}}} +  {{{{\bf{\bar H}}}_k}{{\bf{S}}_k}{\bf{\bar H}}_k^H} {{\left( {{{\bf{H}}_{{\rm{SI}}}}{{\bf{Q}}_k}{\bf{H}}_{{\rm{SI}}}^H + {\sigma ^2}{{\bf{I}}_{{N_r}}}} \right)}^{ - 1}}} \right|}  \label{p1_2_obj}   \\
	&{\rm s.t.}~\eqref{p1_const_3},\eqref{p1_1_const_1},\eqref{p1_1_const_2},\eqref{p1_1_const_3}.
	\end{align}
\end{subequations}
It can be readily checked that   constraint \eqref{p1_1_const_1}  and objective function \eqref{p1_2_obj}  are non-convex, which still makes problem \eqref{p1_2} difficult to solve. To tackle the non-convexity of constraint \eqref{p1_1_const_1},  a local convex approximation is applied. 
The key observation in \eqref{p1_1_const_1} is that the function ${1}/({{1{\rm{ + }}z}})$  is convex with respect to (w.r.t) $z$, and its lower bound can be obtained by taking its first-order Taylor expansion at any feasible point. Specifically, for  any given point ${z^r} = \left\{ {{z_{k,i}^r} \cup z_{k,i}^{0,r},k \in {K_2},i \in {K_2}} \right\}$ in the $r$th iteration, we have
\begin{align}
\frac{1}{{1{\rm{ + }}z}} \ge \frac{1}{{1{\rm{ + }}{z^r}}} - \frac{1}{{{{\left( {1{\rm{ + }}{z^r}} \right)}^2}}}\left( {z - {z^r}} \right)
\overset{\triangle} {=} {f^{{\rm{lb}}}}\left( {z|{z^r}} \right),
\end{align}
which is convex  since ${f^{{\rm{lb}}}}\left( {z|{z^r}} \right)$ is linear with $z$.
As a result, \eqref{p1_1_const_1} can be written as 
\begin{align}
{\tau _k}{\rm{tr}}\left( {{{\bf{S}}_k}} \right) \le \sum\limits_{i = 1}^K {{\tau _{0i}}\left( {{X_k}{f^{{\rm{lb}}}}\left( {z_{k,i}^0|z_{k,i}^{0,r}} \right) - {Y_k}} \right)}  + \sum\limits_{j = 1,j \ne k}^K {{\tau _j}\left( {{X_k}{f^{{\rm{lb}}}}\left( {{z_{k,j}}|z_{k,j}^r} \right) - {Y_k}} \right)} ,k \in {{\cal K}_2}, \label{p1_1_const_1_new}
\end{align}
which is  convex.

We observe that  objective function \eqref{p1_2_obj} is non-convex since  optimization variables ${\bf S}_k$  and  ${\bf Q}_k$ are  coupled with each other  and   ${\bf Q}_k$ is involved with  the inverse operation in the  logarithmic form. To make it more tractable, we rewrite  objective function \eqref{p1_2_obj} as 
\begin{align}
\mathop {\max }\limits_{{\bf{S}}\succeq{\bf{0}},{\bf{Q}}\succeq{\bf{0}},z \ge 0} \sum\limits_{k = 1}^K {{\tau _k}{\omega _k}\ln \left| {{{\bf{H}}_{{\rm{SI}}}}{{\bf{Q}}_k}{\bf{H}}_{{\rm{SI}}}^H + {\sigma ^2}{{\bf{I}}_{{N_r}}} + {{{\bf{\bar H}}}_k}{{\bf{S}}_k}{\bf{\bar H}}_k^H} \right|}  - \sum\limits_{k = 1}^K {{\tau _k}{\omega _k}\ln \left| {{{\bf{H}}_{{\rm{SI}}}}{{\bf{Q}}_k}{\bf{H}}_{{\rm{SI}}}^H + {\sigma ^2}{{\bf{I}}_{{N_r}}}} \right|}.  \label{p1_2_obj_new} 
\end{align}
It can be seen that  optimization variables  ${\bf S}_k$  and  ${\bf Q}_k$ are decoupled in \eqref{p1_2_obj_new}.
In addition, we observe that \eqref{p1_2_obj_new}  has a  difference-of-concave form with two different concave functions, which motivates us to apply the SCA technique to linearize the last concave function into a linear form. Specifically, taking the first-order Taylor expansion
of $\ln \left| {{{\bf{H}}_{{\rm{SI}}}}{{\bf{Q}}_k}{\bf{H}}_{{\rm{SI}}}^H + {\sigma ^2}{{\bf{I}}_{{N_r}}}} \right|$ at any given point ${{\bf{Q}}^r} = \left\{ {{\bf{Q}}_i^r,i \in {K_2}} \right\}$  in the $r$th iteration, its upper bound is  given by  
\begin{align}
&\ln \left| {{{\bf{H}}_{{\rm{SI}}}}{{\bf{Q}}_k}{\bf{H}}_{{\rm{SI}}}^H + {\sigma ^2}{{\bf{I}}_{{N_r}}}} \right| \le \ln \left| {{{\bf{H}}_{{\rm{SI}}}}{\bf{Q}}_k^r{\bf{H}}_{{\rm{SI}}}^H + {\sigma ^2}{{\bf{I}}_{{N_r}}}} \right| + \notag\\
& {\rm{tr}}\left( {{{\left( {{{\bf{H}}_{{\rm{SI}}}}{\bf{Q}}_k^r{\bf{H}}_{{\rm{SI}}}^H + {\sigma ^2}{{\bf{I}}_{{N_r}}}} \right)}^{ - 1}}\left( {{{\bf{H}}_{{\rm{SI}}}}{{\bf{Q}}_k}{\bf{H}}_{{\rm{SI}}}^H - {{\bf{H}}_{{\rm{SI}}}}{\bf{Q}}_k^r{\bf{H}}_{{\rm{SI}}}^H} \right)} \right) \overset{\triangle} {= }{f^{{\rm{up}}}}\left( {{{\bf{Q}}_k}|{\bf{Q}}_k^r} \right),
\end{align}
which is convex since ${f^{{\rm{up}}}}\left( {{{\bf{Q}}_k}|{\bf{Q}}_k^r} \right)$ is linear with ${{{\bf{Q}}_k}}$.

Furthermore, although constraints \eqref{p1_1_const_2} and \eqref{p1_1_const_3} are convex, we  perform some algebra operations on them to obtain a more  simple form given by  
\begin{align}
&{\rm{tr}}\left( {{\bf{H}}_{k,i}^0{\bf{Q}}_i^0{\bf{H}}_{k,i}^{0,H}} \right) \ge {b_k} - \frac{{\ln z_{k,i}^0}}{{{a_k}}},k \in {{\cal K}_2},i \in {{\cal K}_2},\label{p1_1_const_2_new}\\
& {\rm{tr}}\left( {{{\bf{H}}_{k,j}}{{\bf{Q}}_j}{\bf{H}}_{k,j}^H} \right) \ge {b_k} - \frac{{\ln {z_{k,j}}}}{{{a_k}}},k\in {{\cal K}_2},j\in {{\cal K}_2}.\label{p1_1_const_3_new}
\end{align}
It can be seen that both constraints \eqref{p1_1_const_2_new} and \eqref{p1_1_const_3_new} are also convex, and  the left-hand sides of  them are  linear with  the corresponding optimization variables.

As a result, for any given local points $\{{\bf{Q}}^r,z^r\}$, we obtain  the following optimization problem
\begin{subequations}\label{p1_2_new}
\begin{align}
&\mathop {\max }\limits_{{\bf{S}}\succeq{\bf{0}},{\bf{Q}}\succeq{\bf{0}},z \ge 0} \sum\limits_{k = 1}^K {{\tau _k}{\omega _k}\ln \left| {{{\bf{H}}_{{\rm{SI}}}}{{\bf{Q}}_k}{\bf{H}}_{{\rm{SI}}}^H + {\sigma ^2}{{\bf{I}}_{{N_r}}} + {{{\bf{\bar H}}}_k}{{\bf{S}}_k}{\bf{\bar H}}_k^H} \right|}  - \sum\limits_{k = 1}^K {{\tau _k}{\omega _k}{f^{{\rm{up}}}}\left( {{{\bf{Q}}_k}|{\bf{Q}}_k^r} \right)}\\
&{\rm s.t.}~\eqref{p1_const_3}, \eqref{p1_1_const_1_new}, \eqref{p1_1_const_2_new}, \eqref{p1_1_const_3_new}.
\end{align}
\end{subequations}
Based on the previous discussions, problem \eqref{p1_2_new} is convex, which can be solved by the standard convex techniques such as the interior-point method \cite{boyd2004convex}. In addition, it readily follows that
the objective value of problem \eqref{p1_2_new} provides  a lower bound for that of problem \eqref{p1_2}.
\subsection{ DL/UL Time Allocation}
For any given  precoding matices and transmit covariance matrices  
$\{{\bf{ S}}, {\bf Q}, z\}$  and IRS phase shifts $\bf{v }$,  the  DL/UL time allocation optimization from   problem \eqref{p1_1} can be simplified as 
\begin{subequations} \label{p1_3}
	\begin{align}
	&\mathop {\max }\limits_{\tau \ge 0 } \sum\limits_{k = 1}^K {{\tau _k}{\omega _k}\ln \left| {{{\bf{I}}_{{N_r}}} +  {{{{\bf{\bar H}}}_k}{{\bf{S}}_k}{\bf{\bar H}}_k^H} {{\left( {{{\bf{H}}_{{\rm{SI}}}}{{\bf{Q}}_k}{\bf{H}}_{{\rm{SI}}}^H + {\sigma ^2}{{\bf{I}}_{{N_r}}}} \right)}^{ - 1}}} \right|}    \\
	&{\rm s.t.}~\eqref{p1_const_2}, \eqref{p1_1_const_1}.
	\end{align}
\end{subequations}
It can be readily seen that problem \eqref{p1_3} is a linear optimization problem, which can be solved by the standard convex optimization
techniques \cite{boyd2004convex}.
\subsection{IRS Phase Shift Optimization}
For any given  precoding matices and transmit covariance matrices $\{{\bf{ S}}, {\bf Q}, z\}$  and DL/UL time allocation $\tau$, the IRS phase shifts in  problem \eqref{p1_1} can be simplified as 
\begin{subequations} \label{p1_4}
	\begin{align}
	&\mathop {\max }\limits_{{\bf{v }}} \sum\limits_{k = 1}^K {{\tau _k}{\omega _k}\ln \left| {{{\bf{I}}_{{N_r}}} +  {{{{\bf{\bar H}}}_k}{{\bf{S}}_k}{\bf{\bar H}}_k^H} {{\left( {{{\bf{H}}_{{\rm{SI}}}}{{\bf{Q}}_k}{\bf{H}}_{{\rm{SI}}}^H + {\sigma ^2}{{\bf{I}}_{{N_r}}}} \right)}^{ - 1}}} \right|}  \label{p1_4_obj}  \\
	&{\rm s.t.}~\eqref{p1_const_4}, \eqref{p1_1_const_2_new},\eqref{p1_1_const_3_new}.
	\end{align}
\end{subequations}
Problem \eqref{p1_4} is a non-convex optimization
problem since objective function \eqref{p1_4_obj},  constraint \eqref{p1_1_const_2_new},  and constraint \eqref{p1_1_const_3_new} can be shown to be
non-concave w.r.t.  phase-shift vectors $\bf v$, and the unit-modulus
constraint on each reflection coefficient  in \eqref{p1_const_4} is also
non-convex. To solve this problem, an EW algorithm is proposed, where the main idea behind it is to optimize one phase-shift element  with the other phase-shift elements  are fixed. Specifically, we first relax   unit-modulus constraint \eqref{p1_const_4} into a convex form given by 
\begin{align}
\left| {{{ {{{{v}}_{i,m}^0}} }}} \right| \le 1,\left| {{{ {{{{v}}_{i,m}}} }}} \right| \le 1, m \in {\cal M}, i   \in {{\cal K}_2}\label{p1_const_4_new}.
\end{align}
Since  objective function \eqref{p1_4_obj},  constraint \eqref{p1_1_const_2_new}, and constraint \eqref{p1_1_const_3_new} are implicit over each phase shift,  the challenge is how to derive the explicit function of each phase shift such that it can be optimized efficiently. 
Denote ${\bf{\bar G}} = \left[ {{{{\bf{\bar g}}}_1}, \ldots ,{{{\bf{\bar g}}}_M}} \right]$ and ${{{\bf{\bar H}}}_{r,k}} = {\left[ {{{{\bf{\bar h}}}_{r,k,1}}, \ldots ,{{{\bf{\bar h}}}_{r,k,M}}} \right]^H}$, where ${{{{\bf{\bar g}}}_m}}\in {\mathbb C}^{N_r\times 1}$ denotes the $m$th column of ${{\bf{\bar G}}}$ and ${{\bf{\bar h}}_{r,k,m}^H}\in {\mathbb C}^{1\times N_d}$ denotes the $m$th row  of ${{{\bf{\bar H}}}_{r,k}}$. As such, the effective UL MIMO channel ${{{\bf{\bar H}}}_k}$ can be rewritten as 
\begin{align}
{{{\bf{\bar H}}}_k} = {{{\bf{\bar H}}}_{d,k}} + \sum\limits_{m = 1}^M {{v_{k,m}}} {{{\bf{\bar g}}}_m}{\bf{\bar h}}_{r,k,m}^H.
\end{align}
Then, we can rewrite ${{{\bf{\bar H}}}_k}{{{\bf{ S}}}_k}{\bf{\bar H}}_k^H$  as 
\begin{align}
{{{\bf{\bar H}}}_k}{{\bf{S}}_k}{\bf{\bar H}}_k^H& = {{{\bf{\tilde H}}}_{k,\bar m}}{{\bf{S}}_k}{\bf{\tilde H}}_{k,\bar m}^H + {{{\bf{\tilde H}}}_{k,\bar m}}{{\bf{S}}_k}v_{k,m}^*{{{\bf{\bar h}}}_{r,k,m}}{\bf{\bar g}}_m^H + {v_{k,m}}{{{\bf{\bar g}}}_m}{\bf{\bar h}}_{r,k,m}^H{{\bf{S}}_k}{\bf{\tilde H}}_{k,\bar m}^H \notag\\
&+ {{{\bf{\bar g}}}_m}{\bf{\bar h}}_{r,k,m}^H{{\bf{S}}_k}{{{\bf{\bar h}}}_{r,k,m}}{\bf{\bar g}}_m^H\overset{\triangle} {=} {\bf \bar F}\left( {{{v}}_{k,m}} \right),
\end{align}
where ${{{\bf{\tilde H}}}_{k,\bar m}} = \left( {{{{\bf{\bar H}}}_{d,k}} + \sum\limits_{j \ne m}^M {{v_{k,j}}} {{{\bf{\bar g}}}_j}{\bf{\bar h}}_{r,k,j}^H} \right)$.

Similarly, denote ${\bf{G}} = {\left[ {{{\bf{g}}_1}, \ldots ,{{\bf{g}}_M}} \right]^H}$ and ${{\bf{H}}_{r,k}} = \left[ {{{\bf{h}}_{r,k,1}}, \ldots ,{{\bf{h}}_{r,k,M}}} \right]$, where ${\bf{g}}_m^H\in {\mathbb C}^{1\times N_t}$ denotes the $m$th row of ${{\bf{ G}}}$ and ${{\bf{ h}}_{r,k,m}}\in {\mathbb C}^{N_d\times 1}$ denotes the $m$th column of ${{{\bf{ H}}}_{r,k}}$. As such, the effective DL MIMO channels ${{\bf{H}}_{k,j}}$ and ${\bf{H}}_{k,i}^0$ can be  respectively rewritten as 
\begin{align}
&{{\bf{H}}_{k,j}}{\rm{ = }}{{\bf{H}}_{d,k}} + \sum\limits_{m = 1}^M {{v_{j,m}}} {{\bf{h}}_{r,k,m}}{\bf{g}}_m^H, \\
&{\bf{H}}_{k,i}^0{\rm{ = }}{{\bf{H}}_{d,k}} + \sum\limits_{m = 1}^M {v_{i,m}^0} {{\bf{h}}_{r,k,m}}{\bf{g}}_m^H.
\end{align}
Thus, we can respectively rewrite  ${{\bf{H}}_{k,j}}{{\bf{Q}}_j}{\bf{H}}_{k,j}^H$ and  ${{\bf{H}}_{k,i}^0{\bf{Q}}_i^0{\bf{H}}_{k,i}^{0,H}}$ as
\begin{align}
&{{\bf{H}}_{k,j}}{{\bf{Q}}_j}{\bf{H}}_{k,j}^H = {{{\bf{\hat H}}}_{k,\bar m}}{{\bf{Q}}_j}{\bf{\hat H}}_{k,\bar m}^H + {{{\bf{\hat H}}}_{k,\bar m}}{{\bf{Q}}_j}{\left( {{v_{j,m}}{{\bf{h}}_{r,k,m}}{\bf{g}}_m^H} \right)^H} + \left( {{v_{j,m}}{{\bf{h}}_{r,k,m}}{\bf{g}}_m^H} \right){{\bf{Q}}_j}{\bf{\hat H}}_{k,\bar m}^H  \notag\\
&\qquad\qquad\quad+\left( {{{\bf{h}}_{r,k,m}}{\bf{g}}_m^H} \right){{\bf{Q}}_j}{\left( {{{\bf{h}}_{r,k,m}}{\bf{g}}_m^H} \right)^H}\overset{\triangle} {=} {\bf F}\left( {{{v}}_{j,m}} \right),\\
&{\bf{H}}_{k,i}^0{\bf{Q}}_i^0{\bf{H}}_{k,i}^{0,H}  = {\bf{\hat H}}_{k,i,\bar m}^0{\bf{Q}}_i^0{\bf{\hat H}}_{k,i,\bar m}^{0,H} + {\bf{\hat H}}_{k,i,\bar m}^0 {\bf{Q}}_i^0{\left( {v_{i,m}^0{{\bf{h}}_{r,k,m}}{\bf{g}}_m^H} \right)^H} + \left( {v_{i,m}^0{{\bf{h}}_{r,k,m}}{\bf{g}}_m^H} \right){\bf{Q}}_i^0{\bf{\hat H}}_{k,i,\bar m}^{0,H}\notag\\
& \qquad\qquad\quad+ \left( {{{\bf{h}}_{r,k,m}}{\bf{g}}_m^H} \right){\bf{Q}}_i^0{\left( {{{\bf{h}}_{r,k,m}}{\bf{g}}_m^H} \right)^H}\overset{\triangle} {=} {\bf F}^0\left( {{{v}}_{i,m}^0} \right),
\end{align}
where ${{{\bf{\hat H}}}_{k,j,\bar m}} = {{\bf{H}}_{d,k}} + \sum\limits_{i \ne m}^M {{v_{j,i}}} {{\bf{h}}_{r,k,i}}{\bf{g}}_i^H$ and ${\bf{\hat H}}_{k,i,\bar m}^0 = {{\bf{H}}_{d,k}} + \sum\limits_{j \ne m}^M {v_{i,j}^0} {{\bf{h}}_{r,k,j}}{\bf{g}}_j^H$.

In addition, we observe that optimizing blocks $\{{\bf v}_i^0\} $ and $\{{\bf v}_i\} $ are significantly different since block $\{{\bf v}_i\} $ appears in the  objective function and constraint, while $\{{\bf v}_i^0\} $ only involves in the  constraint. Therefore,  different  methods  are required. To proceed, we solve above two blocks separately in the following.
\subsubsection{Optimize ${{{v}}_{i,m}^0}$ with the given $\left\{ {{{\bf{v}}_j}} \right\}$ and $\left\{ {{{ {{{v}}_{j,x}^0} }},j \ne i,x \ne m} \right\}$} This subproblem can be expressed as 
\begin{subequations} \label{p1_4_1}
	\begin{align}
	&{\rm Find}{\kern 1pt} {\kern 1pt} {\kern 1pt} {\kern 1pt} {\kern 1pt} {\kern 1pt} {\kern 1pt} {\kern 1pt} {{{{v}}_{i,m}^0} } \label{p1_4_1_obj}  \\
	&{\rm s.t.} ~{\rm{tr}}\left( {{\bf{F}}^0\left( {{{v}}_{i,m}^0} \right)} \right) \ge {b_k} - \frac{{\ln z_{k,i}^0}}{{{a_k}}},k \in {{\cal K}_2}, \label{p1_4_1_const1}\\
	&\quad~ \left| {{{\bf{v}}_{i,m}^0}} \right| \le 1. \label{p1_4_1_const2}
	\end{align}
\end{subequations}
It is not difficult to check that all the constraints in  problem \eqref{p1_4_1} are  convex, while  there is no explicit objective function in  problem \eqref{p1_4_1}. 
To achieve a better converged solution, we further transform
problem \eqref{p1_4_1} into an optimization problem with an explicit
objective function to obtain a   more efficient phase shift solution
to increase the harvested energy by devices. Specifically,  by introducing an auxiliary non-negative optimization  variable $\mu$, problem \eqref{p1_4_1}  is transformed into the following problem 
\begin{subequations} \label{p1_4_1_new}
	\begin{align}
	&\mathop {\max }\limits_{{{\bf{v}}_{i,m}^0}, \mu  \ge 0}~ \mu  \label{p1_4_1_obj_new}  \\
	&{\rm s.t.} ~{\rm{tr}}\left( {{\bf{F}}^0\left( {{{v}}_{i,m}^0} \right)} \right) \ge {b_k} - \frac{{\ln z_{k,i}^0}}{{{a_k}}}+\mu,k \in {{\cal K}_2},\\
	&\qquad \eqref{p1_4_1_const2},
	\end{align}
\end{subequations}
where the  auxiliary variable $\mu$ can be interpreted as the \textit{DL energy harvesting residual}  for IRS phase shift optimization. It can be  readily checked that problem \eqref{p1_4_1_new} is convex, which can be solved by the  standard convex optimization
techniques.

\subsubsection{Optimize ${{{{{v}}_{k,m}}} }$ with the given  $\left\{ {{\bf{v}}_i^0} \right\}$ and $\{ { {{{{v}}_{i,j}}}},i \ne k,j \ne m\} $} This subproblem is given by 
\begin{subequations} \label{p1_4_2}
	\begin{align}
	&\mathop {\max }\limits_{{{\bf{v}}_{k,m}}} \sum\limits_{k = 1}^K {{\tau _k}{\omega _k}\ln \left| {{{\bf{I}}_{{N_r}}} + {\bf{\bar F}}\left( {{{{v}}_{k,m}}} \right){{\left( {{{\bf{H}}_{{\rm{SI}}}}{{\bf{Q}}_k}{\bf{H}}_{{\rm{SI}}}^H + {\sigma ^2}{{\bf{I}}_{{N_r}}}} \right)}^{ - 1}}} \right|}  \\
	&{\rm s.t.} ~{\rm{tr}}\left( {{\bf{F}}\left( {{{{v}}_{k,m}}} \right)} \right) \ge {b_i} - \frac{{\ln {z_{i,k}}}}{{{a_i}}},i \in {{\cal K}_2},\\
	&\quad~ \left| {{{\bf{v}}_{k,m}}} \right| \le 1.
	\end{align}
\end{subequations}
It is not difficult to check that both the objective function and constraints are all convex, which can be solved by the standard convex techniques.
 \begin{algorithm}[!t]
	\caption{EW-based   algorithm  for solving problem  \eqref{p1_1}.}	\label{alg1}
	\begin{algorithmic}[1]
		\STATE  \textbf{Initialize} IRS phase-shift vector  ${{\bf{v}}^r}$,  time allocation $\tau^r$,  $z^r$,  transmit covariance matrices $\{{\bf Q }_k^r\} $,   iteration  index $r=0$, and  threshold $\varepsilon$.
		\STATE  \textbf{repeat}
		\STATE  \quad Update  precoding  matrices and transmit covariance matrices  by solving  problem \eqref{p1_2_new}.
		\STATE  \quad Update   time allocation  by solving  problem  \eqref{p1_3}. 
		\STATE  \quad  \textbf{loop for $k$ and $m$} 
		        \STATE  \quad \quad  Update IRS phase shift  ${ v}_{k,m}^0$  by solving  problem   \eqref{p1_4_1_new}. 
		       \STATE  \quad \quad  Update IRS phase shift  ${ v}_{k,m}$  by solving  problem   \eqref{p1_4_2}.   
		\STATE  \quad    \textbf{end} 
%				\STATE  \quad  \textbf{loop}
%		\STATE  \quad \quad      
%		\STATE  \quad  \textbf{end}
		\STATE \textbf{until}   the fractional increase of the objective value of problem \eqref{p1_1} is below  $\varepsilon$.
		\STATE Reconstruct IRS phase shift  ${{v}}_{k,m}^{\rm 0, opt} = \frac{{{{{v}}_{k,m}^0}}}{{\left| {{{{v}}_{k,m}^0}} \right|}}$ and ${{v}}_{k,m}^{\rm  opt} = \frac{{{{{v}}_{k,m}}}}{{\left| {{{{v}}_{k,m}}} \right|}}$ for $k \in {\cal K}_2,m\in{\cal M}$.
		\STATE Update $\{ {\bf{S}},{\bf{Q}}, \tau \}$  based on the newly  optimized  IRS phase shifts. 
%		\STATE {\bf Output:}  $\{ {\bf{S}},{\bf{Q}}, \tau, {\bf v}\} $.
	\end{algorithmic}
\end{algorithm}
\subsection{Overall Algorithm and  Computational Complexity Analysis} \label{sectionIII_D}
Based on the solutions to the above subproblems, an EW-based algorithm is proposed, which is summarized in Algorithm~\ref{alg1}. Note that since from steps $2$ to  $9$ in Algorithm~\ref{alg1}, the  norm of  phase shifts cannot  be  guaranteed to   one. Thus, we need to normalize the amplitudes of phase shifts to be one in step $10$ and recompute the block  $\{ {\bf{S}},{\bf{Q}}, \tau \}$ based on the normalized phase shifts. It can be readily seen that the objective value of problem \eqref{p1_1}  is non-decreasing over  iterations since each subproblem is solved locally and/or optimally. In addition,  due to the limited HAP transmit power, the objective value of problem \eqref{p1_1} is constrained in a finite value. Therefore, the proposed EW-based algorithm is guaranteed to converge. 

The mainly computational complexity of Algorithm~\ref{alg1} lies from steps $2$ to  $9$.  Specifically, in step $3$, problem \eqref{p1_2_new} has ${2{K^2}{\rm{ + }}3K}$ variables. The first constraint of problem \eqref{p1_2_new} has $2K$ constraints, each of which has $N_d^2$ dimensions. Similarly, the second  constraint of \eqref{p1_2_new} has $K$ constraints, each of which has $N_d^2+1$ dimensions; the third  constraint has $K^2$ constraints, each of which has $N_d^2+1$ dimensions; the fourth constraint has the same number of constraints and dimensions with the third constraint. As  such, the complexity for solving problem \eqref{p1_2_new} in the worst case   is ${\cal O}{\left( {\left( {2{K^2}{\rm{ + }}3K} \right)\left( {2KN_t^2 + \left( {{K^2} + K} \right)\left( {N_d^2 + 1} \right)} \right)} \right)^{3.5}}$ \cite{ben2001lectures}.
 In step $4$, problem  \eqref{p1_3} is a linear optimization problem, 
 whose complexity is ${\cal O}\left( \sqrt{2K} \right)$ \cite{gondzio1996computational}, where $2K$ denotes the number of variables. In step $6$, problem   \eqref{p1_4_1_new} is also a linear optimization problem, whose complexity is ${\cal O}\left( \sqrt{2} \right)$.  In step $7$, problem   \eqref{p1_4_2} involves the logarithmic form, whose 
complexity analysis is similar to step $3$, which is given by ${\cal O}{\left( {2\left( {K + 1} \right)} \right)^{3.5}}$.
Therefore, the  complexity of Algorithm~\ref{alg1} is given by ${\cal O}\left( {L\left( {\left( {2{K^2}{\rm{ + }}3K} \right){{\left( {2KN_t^2 + \left( {{K^2} + K} \right)\left( {N_d^2 + 1} \right)} \right)}^{3.5}} + \sqrt {2K}  + KM\left( {\sqrt 2  + {{\left( {2\left( {K + 1} \right)} \right)}^{3.5}}} \right)} \right)} \right)$, \\
\noindent where $L$ represents the total number of iterations required to reach convergence. The overall complexity of Algorithm~\ref{alg1}  may be practically high when  the number of loop iterations, i.e.,  $KM$,  required in steps $6$ and $7$ is large. To achieve an efficient algorithm with low complexity, we propose an MMSE-based algorithm in the next section.
\section{MMSE-based Algorithm}\label{sectionIV}
In this section, we propose an efficient MMSE-based algorithm by equivalently transforming the original  problem into a more tractable form.  
Specifically, the achievable rate  can be viewed as a data rate for a hypothetical communication system where  user $k$ estimates  the desired signal ${\bf s}_k$ from \eqref{up_receiver_new}  with an estimator ${{\bf{U}}_k} \in {{\mathbb C}^{{N_r} \times N_d}}$. Thus, the estimated signal of ${\bf s}_k$ is given by
%\footnote{Note that there is a slight different from the previous work \cite{shi2011iteratively}, where it estimates the information signal directly, while we estimate the information-carrying signal consisting of beamforming vectors  here. However, it can be readily verified that this will not change the  equivalence after transformations.}
\begin{align}
{{{\bf{\hat s}}}_k} = {\bf{U}}_k^H{{\bf{\hat y}}_k^r}. \label{Pconst4}
\end{align}
As such, the  mean-square error (MSE) matrix is given by
\begin{align}
&{\bf{E}}_k^{{\rm{mse}}} = {\mathbb E}\left\{ {\left( {{{{\bf{\hat s}}}_k} - {{\bf{s}}_k}} \right){{\left( {{{{\bf{\hat s}}}_k} - {{\bf{s}}_k}} \right)}^H}} \right\}\\
& = {\bf{U}}_k^H\left( {{{{\bf{\bar H}}}_k}{{\bf{W}}_k}{\bf{W}}_k^H{\bf{\bar H}}_k^H + {{\bf{H}}_{{\rm{SI}}}}{{\bf{Q}}_k}{\bf{H}}_{{\rm{SI}}}^H + {\sigma ^2}{{\bf{I}}_{{N_r}}}} \right){{\bf{U}}_k} - {\bf{U}}_k^H{{{\bf{\bar H}}}_k}{{\bf{W}}_k} - {\bf{W}}_k^H{\bf{\bar H}}_k^H{{\bf{U}}_k} + {{\bf{I}}_{{N_d}}}.
\end{align}
Following  Theorem $1$ of \cite{shi2011iteratively} and introducing additional variables ${{\bf{F}}_k} \succeq {\bf 0} \in {{\mathbb C}^{N_d \times N_d}}$,   problem \eqref{p1} can be equivalently transformed as 
\begin{subequations} \label{p2}
	\begin{align}
	&\mathop {\max }\limits_{{\bf{W}},{\bf{Q}}\succeq{\bf{0}},{\bf{F}}\succeq{\bf{0}},{\bf{U}},{\bf{v }},\tau \ge 0} \sum\limits_{k = 1}^K {{\tau _k}{\omega _k}\left( {\ln \left| {{{\bf{F}}_k}} \right|  - {\rm{tr}}\left( {{{\bf{F}}_k}{\bf{E}}_k^{{\rm{mse}}}} \right) + {N_d}} \right)} \label{p2_obj} \\
	&{\rm s.t.}~\eqref{p1_const_1},\eqref{p1_const_2},\eqref{p1_const_3},\eqref{p1_const_4},
	\end{align}
\end{subequations}
where  ${\bf{F}} = \left\{ {{{\bf{F}}_i},i \in {{\cal K}_2}} \right\}$ and ${\bf{U}} = \left\{ {{{\bf{U}}_i},i \in {{\cal K}_2}} \right\}$. Although problem \eqref{p2} introduces additional optimization variables $\bf F$ and $\bf U$, 
the structure of the newly formulated  problem  facilitates the design of a computationally efficient suboptimal algorithm.
To begin with, similar to Section~\ref{sectionIII},  we also introduce   auxiliary variables $z$. As such, 
 problem  \eqref{p2} can be rewritten as  
\begin{subequations} \label{p3}
	\begin{align}
	&\mathop {\max }\limits_{{\bf{W}},{\bf{Q}}\succeq{\bf{0}},{\bf{F}}\succeq{\bf{0}},{\bf{U}},{\bf{v }},\tau \ge 0,z\ge 0} \sum\limits_{k = 1}^K {{\tau _k}{\omega _k}\left( {\ln \left| {{{\bf{F}}_k}} \right|- {\rm{tr}}\left( {{{\bf{F}}_k}{\bf{E}}_k^{{\rm{mse}}}} \right) + {N_d}} \right)}\label{p3_obj}\\
	&{\rm s.t.}~\eqref{p1_const_2},\eqref{p1_const_3},\eqref{p1_const_4},\eqref{p1_1_const_1},\eqref{p1_1_const_2},\eqref{p1_1_const_3}.
	\end{align}
\end{subequations}
Then,  we divide all the variables into five blocks, namely  estimators ${\bf U}$, auxiliary variables ${\bf F}$, precoding matrices and transmit covariance matrices $\{{\bf W},{\bf Q},z\}$,  time allocation $\tau$, and IRS phase shifts $\bf v$, and then alternately solve these blocks. 
\subsection{Optimization of  Estimators ${\bf U}$} 
It can be observed from problem  \eqref{p3} that   estimator ${\bf U}_k$ only exists  in  \eqref{p3_obj}. As such, with the given blocks ${\bf F}$,  $\{{\bf W},{\bf Q},z\}$, $\tau$, and $\bf v$, the optimal $\bf U$ can be obtained by taking the first-order derivative of ${\bf E}_k^{\rm mse}$ and setting it zero.  Accordingly, we obtain the optimal solution as 
\begin{align}
{\bf{U}}_k^{{\rm{opt}}}{\rm{ = }}{\left( {{{{\bf{\bar H}}}_k}{{\bf{W}}_k}{{\bf{W}}_k^H}{\bf{\bar H}}_k^H + {{\bf{H}}_{{\rm{SI}}}}{{\bf{Q}}_k}{\bf{H}}_{{\rm{SI}}}^H + {\sigma ^2}{{\bf{I}}_{{N_r}}}} \right)^{{\rm{ - }}1}}{{{\bf{\bar H}}}_k}{{\bf{W}}_k}.\label{optimalu}
\end{align}
\subsection{Optimization of Auxiliary Variables $\bf F$} 
We  observe that ${\bf F}_k$ only appears   in  \eqref{p3_obj}. As such, with the given blocks ${\bf U}$,  $\{{\bf W},{\bf Q},z\}$, $\tau$, and $\bf v$,  the optimal $\bf F$ can be obtained by taking the first-order derivative of \eqref{p3_obj} and setting it zero, which results in the solution given as 
\begin{align}
{\bf{F}}_k^{{\rm{opt}}}{\rm{ = }}{\left( {{\bf{E}}_k^{{\rm{mse}}}} \right)^{{\rm{ - }}1}}.\label{optimalf}
\end{align}
\subsection{Optimization of Precoding  Matrices and Transmit Covariance Matrices $\{{\bf W},{\bf { Q}},z\}$} 
For the  given  blocks ${\bf U}$, ${\bf F}$, $\tau$,  and $\bf{v }$,  the subproblem regarding to precoding matrix and transmit covariance matrix 
optimization is given by 
\begin{subequations} \label{p3_3}
	\begin{align}
	&\mathop {\max }\limits_{{\bf{W}},{\bf{Q}}\succeq{\bf{0}},z\ge 0} \sum\limits_{k = 1}^K {{\tau _k}{\omega _k}\left( {\ln \left| {{{\bf{F}}_k}} \right|  - {\rm{tr}}\left( {{{\bf{F}}_k}{\bf{E}}_k^{{\rm{mse}}}} \right) + {N_d}} \right)} \label{p3_3_obj}\\
	&{\rm s.t.}~\eqref{p1_const_3},\eqref{p1_1_const_1},\eqref{p1_1_const_2},\eqref{p1_1_const_3}.
	\end{align}
\end{subequations}
We observe that problem \eqref{p3_3} has the same  constraints with problem  \eqref{p1_2} in Section~\ref{sectionIII}.  Thus, the constraints in problem \eqref{p3_3}  can be tackled with the same way in problem  \eqref{p1_2} and are omitted here for brevity.
In addition, it is not difficult to see that   objective function \eqref{p3_3_obj} is jointly concave w.r.t. ${\bf W}$ and ${\bf Q}$ owing  to the MMSE-based transformation for decoupling them. Therefore, problem \eqref{p3_3} can be well addressed by the standard convex techniques.
\subsection{Optimization of Time Allocation  $\tau$} 
For the  given  blocks ${\bf U}$, ${\bf F}$, $\{{\bf W},{\bf Q},z\}$,  and $\bf{v }$,  the subproblem regarding to time allocation 
optimization is given by
\begin{subequations} \label{p3_4}
	\begin{align}
&\mathop {\max }\limits_{\tau\ge 0} \sum\limits_{k = 1}^K {{\tau _k}{\omega _k}\left( {\ln \left| {{{\bf{F}}_k}} \right|  - {\rm{tr}}\left( {{{\bf{F}}_k}{\bf{E}}_k^{{\rm{mse}}}} \right) + {N_d}} \right)}\\
	&{\rm s.t.}~\eqref{p1_const_2},\eqref{p1_1_const_1},
	\end{align}
\end{subequations}
which  is a linear  optimization problem.
\subsection{Optimization of IRS Phase Shifts $\bf v$}
For the  given  blocks ${\bf U}$, ${\bf F}$, $\{{\bf W},{\bf Q},z\}$,  and $\tau$,  the subproblem regarding to  IRS phase shift
optimization is given by
\begin{subequations} \label{p4}
	\begin{align}
	&\mathop {\max }\limits_{\bf v} \sum\limits_{k = 1}^K {{\tau _k}{\omega _k}\left( {\ln \left| {{{\bf{F}}_k}} \right|   - {\rm{tr}}\left( {{{\bf{F}}_k}{\bf{E}}_k^{{\rm{mse}}}} \right) + {N_d}} \right)}\label{p4_obj}\\
	&{\rm s.t.}~ \eqref{p1_const_4},\eqref{p1_1_const_2},\eqref{p1_1_const_3}.
	\end{align}
\end{subequations}
Problem \eqref{p4_obj} is still a non-convex optimization problem. To obtain an efficient solution, 
we first respectively expand  ${\rm{tr}}\left( {{{\bf{F}}_k}{\bf{U}}_k^H{{{\bf{\bar H}}}_k}{{{\bf{W}}}_k}{{{\bf{W}}}_k^H}{\bf{\bar H}}_k^H{{\bf{U}}_k}} \right)$, 
${\rm{tr}}\left( {{{\bf{F}}_k}{\bf{U}}_k^H{{{\bf{\bar H}}}_k}{{{\bf{ W}}}_k}} \right)$, and ${\rm{tr}}\left( {{{\bf{F}}_k}{\bf{W}}_k^H{\bf{\bar H}}_k^H{{\bf{U}}_k}} \right)$ as 
\begin{align}
&{\rm{tr}}\left( {{{\bf{F}}_k}{\bf{U}}_k^H{{{\bf{\bar H}}}_k}{{{\bf{ W}}}_k}{{{\bf{ W}}}_k^H}{\bf{\bar H}}_k^H{{\bf{U}}_k}} \right)= {\rm{tr}}\left( {{{{\bf{\bar H}}}_{d,k}}{{{\bf{ W}}}_k}{{{\bf{ W}}}_k^H}{\bf{\bar H}}_{d,k}^H{{\bf{U}}_k}{{\bf{F}}_k}{\bf{U}}_k^H} \right) \notag\\
&  + {\rm{tr}}\left( {{{{\bf{\bar H}}}_{d,k}}{{{\bf{ W}}}_k}{{{\bf{ W}}}_k^H}{\bf{\bar H}}_{r,k}^H{\bf{\Theta }}_k^H{\bf{\bar G}}^H{{\bf{U}}_k}{{\bf{F}}_k}{\bf{U}}_k^H} \right)+{\rm{tr}}\left( {{\bf{\bar G}}{{\bf{\Theta }}_k}{{{\bf{\bar H}}}_{r,k}}{{{\bf{ W}}}_k}{{{\bf{ W}}}_k^H}{\bf{\bar H}}_{d,k}^H{{\bf{U}}_k}{{\bf{F}}_k}{\bf{U}}_k^H} \right)  \notag\\
& + {\rm{tr}}\left( {{\bf{\bar G}}{{\bf{\Theta }}_k}{{{\bf{\bar H}}}_{r,k}}{{{\bf{ W}}}_k}{{{\bf{ W}}}_k^H}{\bf{\bar H}}_{r,k}^H{\bf{\Theta }}_k^H{\bf{\bar G}}^H{{\bf{U}}_k}{{\bf{F}}_k}{\bf{U}}_k^H} \right),\\
&{\rm{tr}}\left( {{{\bf{F}}_k}{\bf{U}}_k^H{{{\bf{\bar H}}}_k}{{{\bf{ W}}}_k}} \right) = {\rm{tr}}\left( {{{\bf{F}}_k}{\bf{U}}_k^H{{{\bf{\bar H}}}_{d,k}}{{{\bf{ W}}}_k}} \right) + {\rm{tr}}\left( {{{\bf{F}}_k}{\bf{U}}_k^H{\bf{\bar G}}{{\bf{\Theta }}_k}{{{\bf{\bar H}}}_{r,k}}{{{\bf{ W}}}_k}} \right),\\
&{\rm{tr}}\left( {{{\bf{F}}_k}{\bf{W}}_k^H{\bf{\bar H}}_k^H{{\bf{U}}_k}} \right) = {\rm{tr}}\left( {{{\bf{F}}_k}{\bf{ W}}_k^H{\bf{\bar H}}_{d,k}^H{{\bf{U}}_k}} \right) + {\rm{tr}}\left( {{{\bf{F}}_k}{\bf{W}}_k^H{\bf{\bar H}}_{r,k}^H{\bf{\Theta }}_k^H{\bf{\bar G}}^H{{\bf{U}}_k}} \right).
\end{align}
Then, define ${{\bf{c}}_k} = {\left[ {{{\left[ {{{{\bf{\bar H}}}_{r,k}}{{{\bf{ W}}}_k}{{{\bf{ W}}}_k^H}{\bf{\bar H}}_{d,k}^H{{\bf{U}}_k}{{\bf{F}}_k}{\bf{U}}_k^H{\bf{\bar G}}} \right]}_{1,1}}, \ldots ,{{\left[ {{{{\bf{\bar H}}}_{r,k}}{{{\bf{ W}}}_k}{{{\bf{ W}}}_k^H}{\bf{\bar H}}_{d,k}^H{{\bf{U}}_k}{{\bf{F}}_k}{\bf{U}}_k^H{\bf{\bar G}}} \right]}_{M,M}}} \right]^T}$ and  ${{{\bf{\bar c}}}_k} = {\left[ {{{\left[ {{{{\bf{\bar H}}}_{r,k}}{{{\bf{ W}}}_k}{{\bf{F}}_k}{\bf{U}}_k^H{\bf{\bar G}}} \right]}_{1,1}}, \ldots ,{{\left[ {{{{\bf{\bar H}}}_{r,k}}{{{\bf{ W}}}_k}{{\bf{F}}_k}{\bf{U}}_k^H{\bf{\bar G}}} \right]}_{M,M}}} \right]^T}$,
we have 
\begin{align}
&{\rm{tr}}\left( {{{{\bf{\bar H}}}_{d,k}}{{{\bf{ W}}}_k}{{{\bf{ W}}}_k^H}{\bf{\bar H}}_{r,k}^H{\bf{\Theta }}_k^H{\bf{\bar G}}^H{{\bf{U}}_k}{{\bf{F}}_k}{\bf{U}}_k^H} \right) \!\!=\!\! {\rm{tr}}\left( {{\bf{\Theta }}_k^H{\bf{\bar G}}^H{{\bf{U}}_k}{{\bf{F}}_k}{\bf{U}}_k^H{{{\bf{\bar H}}}_{d,k}}{{{\bf{ W}}}_k}{{{\bf{ W}}}_k^H}{\bf{\bar H}}_{r,k}^H} \right) \!\!=\!\! {\bf{c}}_k^H{\bf{v}}_k^*,\\
&{\rm{tr}}\left( {{\bf{\bar G}}{{\bf{\Theta }}_k}{{{\bf{\bar H}}}_{r,k}}{{{\bf{ W}}}_k}{{{\bf{ W}}}_k^H}{\bf{\bar H}}_{d,k}^H{{\bf{U}}_k}{{\bf{F}}_k}{\bf{U}}_k^H} \right) = {\rm{tr}}\left( {{{\bf{\Theta }}_k}{{{\bf{\bar H}}}_{r,k}}{{{\bf{ W}}}_k}{{{\bf{ W}}}_k^H}{\bf{\bar H}}_{d,k}^H{{\bf{U}}_k}{{\bf{F}}_k}{\bf{U}}_k^H{\bf{\bar G}}} \right) = {\bf{v}}_k^T{{\bf{c}}_k},\\
&{\rm{tr}}\left( {{\bf{\bar G}}{{\bf{\Theta }}_k}{{{\bf{\bar H}}}_{r,k}}{{{\bf{ W}}}_k}{{{\bf{ W}}}_k^H}{\bf{\bar H}}_{r,k}^H{\bf{\Theta }}_k^H{{{\bf{\bar G}}}^H}{{\bf{U}}_k}{{\bf{F}}_k}{\bf{U}}_k^H} \right) = {\bf{v}}_k^H{{{\bf{\bar D}}}_k}{{\bf{v}}_k},\\
&{\rm{tr}}\left( {{{\bf{F}}_k}{\bf{U}}_k^H{\bf{\bar G}}{{\bf{\Theta }}_k}{{{\bf{\bar H}}}_{r,k}}{{{\bf{ W}}}_k}} \right) = {\rm{tr}}\left( {{{\bf{\Theta }}_k}{{{\bf{\bar H}}}_{r,k}}{{{\bf{W}}}_k}{{\bf{F}}_k}{\bf{U}}_k^H{\bf{\bar G}}} \right) = {\bf{v}}_k^T{{{\bf{\bar c}}}_k},\\
&{\rm{tr}}\left( {{{\bf{F}}_k}{\bf{ W}}_k^H{\bf{\bar H}}_{r,k}^H{\bf{\Theta }}_k^H{ \bf{\bar G}}^H{{\bf{U}}_k}} \right) = {\bf{\bar c}}_k^H{\bf{v}}_k^*,
\end{align}
where ${{{\bf{\bar D}}}_k} = \left( {{{{\bf{\bar G}}}^H}{{\bf{U}}_k}{{\bf{F}}_k}{\bf{U}}_k^H{\bf{\bar G}}} \right) \odot {\left( {{{{\bf{\bar H}}}_{r,k}}{{{\bf{ W}}}_k}{{{\bf{ W}}}_k^H}{\bf{\bar H}}_{r,k}^H} \right)^T}$.
As such, we have 
\begin{align}
&{\rm{tr}}\left( {{{\bf{F}}_k}{\bf{U}}_k^H{{{\bf{\bar H}}}_k}{{{\bf{ W}}}_k}{{{\bf{ W}}}_k^H}{\bf{\bar H}}_k^H{{\bf{U}}_k}} \right){\rm{ = tr}}\left( {{{{\bf{\bar H}}}_{d,k}}{{{\bf{ W}}}_k}{{{\bf{ W}}}_k^H}{\bf{\bar H}}_{d,k}^H{{\bf{U}}_k}{{\bf{F}}_k}{\bf{U}}_k^H} \right)+\notag\\
&\qquad\qquad\qquad\qquad\qquad\qquad  + {\bf{v}}_k^T{{\bf{c}}_k} + {\bf{c}}_k^H{\bf{v}}_k^* + {\bf{v}}_k^H{{{\bf{\bar D}}}_k}{{\bf{v}}_k},\\
&{\rm{tr}}\left( {{{\bf{F}}_k}{\bf{U}}_k^H{{{\bf{\bar H}}}_k}{{\bf{W}}_k}} \right) = {\rm{tr}}\left( {{{\bf{F}}_k}{\bf{U}}_k^H{{{\bf{\bar H}}}_{d,k}}{{\bf{W}}_k}} \right) + {\bf{v}}_k^T{{{\bf{\bar c}}}_k},\\
&{\rm{tr}}\left( {{{\bf{F}}_k}{\bf{W}}_k^H{\bf{\bar H}}_k^H{{\bf{U}}_k}} \right) = {\rm{tr}}\left( {{{\bf{F}}_k}{\bf{W}}_k^H{\bf{\bar H}}_{d,k}^H{{\bf{U}}_k}} \right) + {\bf{\bar c}}_k^H{\bf{v}}_k^*.
\end{align}
In addition, for   constraint  \eqref{p1_1_const_3},
 we expand   ${\rm{tr}}\left( {{{\bf{H}}_{k,j}}{{\bf{Q}}_j}{\bf{H}}_{k,j}^H} \right)$ as 
\begin{align}
{\rm{tr}}\left( {{{\bf{H}}_{k,j}}{{\bf{Q}}_j}{\bf{H}}_{k,j}^H} \right) &={\rm{tr}}\left( {{{\bf{H}}_{d,k}}{{\bf{Q}}_j}{\bf{H}}_{d,k}^H} \right){\rm{ + tr}}\left( {{{\bf{H}}_{d,k}}{{\bf{Q}}_j}{{\bf{G}}^H}{\bf{\Theta }}_j^H{\bf{H}}_{r,k}^H} \right)\notag\\
&+{\rm{tr}}\left( {{{\bf{H}}_{r,k}}{{\bf{\Theta }}_j}{\bf{G}}{{\bf{Q}}_j}{\bf{H}}_{d,k}^H} \right){\rm{ + tr}}\left( {{{\bf{H}}_{r,k}}{{\bf{\Theta }}_j}{\bf{G}}{{\bf{Q}}_j}{{\bf{G}}^H}{\bf{\Theta }}_j^H{\bf{H}}_{r,k}^H} \right).
\end{align}
Define   ${\bf z}_{k,j} = {\left[ {{{\left[ {{\bf{G}}{{\bf{Q}}_j}{\bf{H}}_{d,k}^H{{\bf{H}}_{r,k}}} \right]}_{1,1}}, \ldots ,{{\left[ {{\bf{G}}{{\bf{Q}}_j}{\bf{H}}_{d,k}^H{{\bf{H}}_{r,k}}} \right]}_{M,M}}} \right]^T}$, we have
\begin{align}
&{\rm{tr}}\left( {{{\bf{H}}_{r,k}}{{\bf{\Theta }}_j}{\bf{G}}{{\bf{Q}}_j}{\bf{H}}_{d,k}^H} \right) = {\rm{tr}}\left( {{{\bf{\Theta }}_j}{\bf{G}}{{\bf{Q}}_j}{\bf{H}}_{d,k}^H{{\bf{H}}_{r,k}}} \right) = {\bf{v}}_j^T{\bf z}_{k,j},\notag\\
&{\rm{tr}}\left( {{{\bf{H}}_{d,k}}{{\bf{Q}}_j}{{\bf{G}}^H}{\bf{\Theta }}_j^H{\bf{H}}_{r,k}^H} \right) = {{\bf{z}}_{k,j}^H}{\bf{v}}_j^*,\notag\\
&{\rm{tr}}\left( {{{\bf{H}}_{r,k}}{{\bf{\Theta }}_j}{\bf{G}}{{\bf{Q}}_j}{{\bf{G}}^H}{\bf{\Theta }}_j^H{\bf{H}}_{r,k}^H} \right) = {\rm{tr}}\left( {{\bf{\Theta }}_j^H{\bf{H}}_{r,k}^H{{\bf{H}}_{r,k}}{{\bf{\Theta }}_j}{\bf{G}}{{\bf{Q}}_j}{{\bf{G}}^H}} \right) = {\bf{v}}_j^H{{\bf{D}}_{k,j}}{{\bf{v}}_j},
\end{align}
where ${{\bf{D}}_{k,j}} = \left( {{\bf{H}}_{r,k}^H{{\bf{H}}_{r,k}}} \right) \odot {\left( {{\bf{G}}{{\bf{Q}}_j}{{\bf{G}}^H}} \right)^T}$. As such, we can rewrite ${\rm{tr}}\left( {{{\bf{H}}_{k,j}}{{\bf{Q}}_j}{\bf{H}}_{k,j}^H} \right)$ as 
\begin{align}
{\rm{tr}}\left( {{{\bf{H}}_{k,j}}{{\bf{Q}}_j}{\bf{H}}_{k,j}^H} \right){\rm{ = }}{\bf{v}}_j^H{{\bf{D}}_{k,j}}{{\bf{v}}_j} + {\bf{v}}_j^T{{\bf{z}}_{k,j}} + {\bf{z}}_{k,j}^H{\bf{v}}_j^* + {\rm{tr}}\left( {{{\bf{H}}_{d,k}}{{\bf{Q}}_j}{\bf{H}}_{d,k}^H} \right). \label{ULenegy}
\end{align}
Similar to \eqref{ULenegy}, we can rewrite ${\rm{tr}}\left( {{\bf{H}}_{k,i}^0{\bf{Q}}_i^0{\bf{H}}_{k,i}^{0,H}} \right)$  in \eqref{p1_1_const_2} as 
\begin{align}
{\rm{tr}}\left( {{\bf{H}}_{k,i}^0{\bf{Q}}_i^0{\bf{H}}_{k,i}^{0,H}} \right){\rm{ = }}{\bf{v}}_i^{0,H}{\bf{D}}_{k,i}^0{\bf{v}}_i^0 + {\bf{v}}_i^{0,T}{{\bf{z}}_{k,i}^0} + {\bf{z}}_{k,i}^{0,H}{\bf{v}}_0^* + {\rm{tr}}\left( {{{\bf{H}}_{d,k}}{{\bf{Q}}_i^0}{\bf{H}}_{d,k}^H} \right),
\end{align}
with ${\bf{D}}_{k,i}^0 = \left( {{\bf{H}}_{r,k}^H{{\bf{H}}_{r,k}}} \right) \odot {\left( {{\bf{G}}{{\bf{Q}}_i^0}{{\bf{G}}^H}} \right)^T}$ and ${\bf{z}}_{k,i}^0 = {\left[ {{{\left[ {{\bf{GQ}}_i^0{\bf{H}}_{d,k}^H{{\bf{H}}_{r,k}}} \right]}_{1,1}}, \ldots ,{{\left[ {{\bf{GQ}}_i^0{\bf{H}}_{d,k}^H{{\bf{H}}_{r,k}}} \right]}_{M,M}}} \right]^T}$.
Therefore,  problem \eqref{p4} can be transformed as (ignore the irrelevant constants w.r.t. $\bf v$)
\begin{subequations} \label{p4_1}
\begin{align}
&\mathop {\min }\limits_{\bf{v}} \sum\limits_{k = 1}^K {{\tau _k}{\omega _k}} \left( {{\bf{v}}_k^H{{{\bf{\bar D}}}_k}{{\bf{v}}_k} + {\bf{v}}_k^T\left( {{{\bf{c}}_k} - {{{\bf{\bar c}}}_k}} \right) + \left( {{\bf{c}}_k^H - {\bf{\bar c}}_k^H} \right){\bf{v}}_k^*} \right)\label{p4_1_obj}\\
& {\rm s.t.}~{\bf{v}}_j^H{{\bf{D}}_{k,j}}{{\bf{v}}_j} + {\bf{v}}_j^T{{\bf{z}}_{k,j}} + {\bf{z}}_{k,j}^H{\bf{v}}_j^* + {\rm{tr}}\left( {{{\bf{H}}_{d,k}}{{\bf{Q}}_j}{\bf{H}}_{d,k}^H} \right) \ge {b_k} - \frac{{\ln {z_{k,j}}}}{{{a_k}}},k \in {{\cal K}_2},j \in {{\cal K}_2},\label{p4_1_const1}\\
& \qquad {\bf{v}}_i^{0,H}{\bf{D}}_{k,i}^0{\bf{v}}_i^0 + {\bf{v}}_i^{0,T}{{\bf{z}}_{k,i}^0} + {\bf{z}}_{k,i}^{0,H}{\bf{v}}_0^* + {\rm{tr}}\left( {{{\bf{H}}_{d,k}}{{\bf{Q}}_i^0}{\bf{H}}_{d,k}^H} \right) \ge {b_k} - \frac{{\ln z_{k,i}^0}}{{{a_k}}}, k \in {K_2},i \in {K_2},\label{p4_1_const2}\\
&\qquad \eqref{p1_const_4}.
\end{align}
\end{subequations}
It is observed that \eqref{p4_1_obj} is a quadratic function of $\bf v$, which is convex. Although constraints \eqref{p4_1_const1} and \eqref{p4_1_const2} are non-convex,  we observe that the left-hand sides of \eqref{p4_1_const1} and \eqref{p4_1_const2} are quadratic functions of $\bf v$, which motivates us to apply the  SCA technique to tackle them. Specifically,  for any given local points ${\bf v}_j^r$ and ${\bf v}_i^{0,r}$, we  respectively obtain its lower bound given by  
\begin{align}
&{\bf{v}}_j^H{{\bf{D}}_{k,j}}{{\bf{v}}_j} \ge  - {\bf{v}}_j^{r,H}{{\bf{D}}_{k,j}}{\bf{v}}_j^r + 2{\mathop{\rm Re}\nolimits} \left\{ {{\bf{v}}_j^{r,H}{{\bf{D}}_{k,j}}{{\bf{v}}_j}} \right\}\overset{\triangle} {=} {g^{{\rm{lb}}}}\left( {{{\bf{v}}_j}} \right),\label{sca_uplinkphase}\\
&{\bf{v}}_i^{0,H}{\bf{D}}_{k,i}^0{\bf{v}}_i^0 \ge  - {\bf{v}}_i^{0,r,H}{\bf{D}}_{k,i}^0{\bf{v}}_i^{0,r} + 2{\mathop{\rm Re}\nolimits} \left\{ {{\bf{v}}_i^{0,r,H}{\bf{D}}_{k,i}^0{\bf{v}}_i^0} \right\} \overset{\triangle} {=} {{\tilde g}^{{\rm{lb}}}}\left( {{\bf{v}}_i^0} \right),\label{sca_dllinkphase}
\end{align}
which are convex.
In addition, we relax unit-modulus constraint \eqref{p1_const_4} as in 
\eqref{p1_const_4_new}. As a result, with  \eqref{p1_const_4_new}, \eqref{sca_uplinkphase}, and \eqref{sca_dllinkphase}, we have the newly formulated optimization problem given by 
\begin{subequations} \label{p4_1_new}
	\begin{align}
	&\mathop {\min }\limits_{\bf{v}} \sum\limits_{k = 1}^K {{\tau _k}{\omega _k}} \left( {{\bf{v}}_k^H{{{\bf{\bar D}}}_k}{{\bf{v}}_k} + {\bf{v}}_k^T\left( {{{\bf{c}}_k} - {{{\bf{\bar c}}}_k}} \right) + \left( {{\bf{c}}_k^H - {\bf{\bar c}}_k^H} \right){\bf{v}}_k^*} \right)\\
	& {\rm s.t.}~{g^{{\rm{lb}}}}\left( {{{\bf{v}}_j}} \right) + {\bf{v}}_j^T{{\bf{z}}_{k,j}} + {\bf{z}}_{k,j}^H{\bf{v}}_j^* + {\rm{tr}}\left( {{{\bf{H}}_{d,k}}{{\bf{Q}}_j}{\bf{H}}_{d,k}^H} \right) \ge {b_k} - \frac{{\ln {z_{k,j}}}}{{{a_k}}},k \in {{\cal K}_2},j \in {{\cal K}_2},\\
	& \qquad {{\tilde g}^{{\rm{lb}}}}\left( {{\bf{v}}_i^0} \right) + {\bf{v}}_i^{0,T}{{\bf{z}}_{k,i}^0} + {\bf{z}}_{k,i}^{0,H}{\bf{v}}_0^* + {\rm{tr}}\left( {{{\bf{H}}_{d,k}}{{\bf{Q}}_i^0}{\bf{H}}_{d,k}^H} \right) \ge {b_k} - \frac{{\ln z_{k,i}^0}}{{{a_k}}}, k \in {K_2},i \in {K_2},\\
	&\qquad \eqref{p1_const_4_new}.
	\end{align}
\end{subequations}
It is observed that  problem \eqref{p4_1_new} is a  standard  SOCP problem, which thus can be solved by  the standard convex techniques.
 \begin{algorithm}[!t]
	\caption{MMSE-based   algorithm  for solving problem  \eqref{p1}.}	\label{alg2}
	\begin{algorithmic}[1]
		\STATE  \textbf{Initialize} phase shift   ${{\bf{v}}^r}$,  time allocation $\tau^r$,  $z^r$,   iteration  index $r=0$, and  threshold $\varepsilon$.
		\STATE  \textbf{repeat}
		\STATE  \quad Update   estimators  $\bf U$ based on \eqref{optimalu}.
		\STATE  \quad Update   auxiliary variables $\bf F$ based on \eqref{optimalf}. 
		\STATE   \quad  Update precoding matrices and covariance matrices   $\{{\bf W},{\bf Q},z\}$  by solving  problem \eqref{p3_3}.    
				\STATE  \quad  Update time allocation   $\tau$  by solving  problem   \eqref{p3_4}. 
		\STATE  \quad  Update phase shifts   ${\bf v}$  by solving  problem   \eqref{p4_1_new}.    
		\STATE \textbf{until}   the fractional increase of the objective value of problem \eqref{p1} is below $\varepsilon$.
		\STATE Reconstruct  phase shift  ${{v}}_{k,m}^{ 0, \rm opt} = \frac{{{{{v}}_{k,m}^0}}}{{\left| {{{{v}}_{k,m}^0}} \right|}}$ and ${{v}}_{k,m}^{\rm  opt} = \frac{{{{{v}}_{k,m}}}}{{\left| {{{{v}}_{k,m}}} \right|}}$ for $k \in {\cal K}_2,m\in{\cal M}$.
		\STATE Update $\{ {\bf{U}},{\bf{F}},{\bf{W}},{\bf{Q}},z,\tau \} $  based on the newly  optimized  phase shifts. 
	\end{algorithmic}
\end{algorithm}
\subsection{Overall Algorithm and  Computational Complexity Analysis}
Based on the solutions to its above subproblems, we propose an MMSE-based algorithm, which is summarized in Algorithm~\ref{alg2}. The convergence of Algorithm~\ref{alg2} can be theoretically analyzed  from Theorem $3$ in \cite{shi2011iteratively}.

The computational complexity  of   Algorithm~\ref{alg2} is  analyzed   as follows.  In steps $3$ and $4$, the complexity of computing ${\bf U}$ and ${\bf F}$ are respectively given by  ${\cal O}\left( {KN_r^3} \right)$ and  ${\cal O}\left( {KN_d^3} \right)$. In steps $5$ and $6$, the complexity of computing $\{{\bf W},{\bf Q},z\}$ and $\tau $ are the same as in problems \eqref{p1_2_new} and \eqref{p1_3}, respectively,  whose complexity are  given in Section~ \ref{sectionIII_D}. In step $7$,  
problem \eqref{p4_1_new} is an  SOCP problem, which has $2K$ optimization vectors. 
 The first constraint  has   $K^2$ constraints, each of which has $M$ dimensions;  the second constraint has the same number of constraints and dimensions with the first constraint;  the third  constraint has $2KM$ constraints, each of which has $1$ dimension. As such, the complexity for solving problem \eqref{p4_1_new} is given by ${\cal O}{\left( {2K\left( {2{K^2}M + {2KM}} \right)} \right)}$ \cite{lobo1998applications}. Therefore, the overall   complexity of Algorithm~\ref{alg2} is given by ${\cal O}\Big({L_{{\rm{iter}}}}\Big({\left( {\left( {2{K^2}{\rm{ + }}3K} \right)\left( {2KN_t^2 + \left( {{K^2} + K} \right)\left( {N_d^2 + 1} \right)} \right)} \right)^{3.5}} + \sqrt {2K}  + 2K\left( {2{K^2}M + 2KM} \right) + K$  $\left( {N_r^3 + N_d^3} \right)\Big)\Big)$, where $L_{\rm  iter}$ represents the total number of iterations required to reach convergence. Obviously, compared to the EW-based Algorithm~\ref{alg1}, the complexity of the MMSE-based Algorithm~\ref{alg2} is significantly reduced.

 \section{Numerical Results}
 In this section, we provide numerical results   to verify  the effectiveness of the proposed algorithms  and to provide useful insights for the IRS-aided MIMO FD-WPCN. 
% We consider a system that operates on a carrier frequency of $750~{\rm MHz}$ with the system bandwidth of $1~{\rm MHz}$ and   effective noise power density   $-150~{\rm  dBm/Hz}$ \cite{wu2020jointActive}. We assume that the  IRS  is equipped with  a uniform rectangular array with $M=M_xM_z$ reflecting elements, where $M_x$ and $M_z$ denotes the numbers of reflecting elements along the $x$-axis and $z$-axis, respectively. We fix $M_x=5$ and increase $M_z$ linearly with $M$.  We assume that the antenna spacing is half of the wavelength, i.e., $\lambda /2 = 0.2~\rm m$.  
A three-dimensional   coordinate setup measured in meter (m) is considered, where the HAP and the IRS are located at $( 0, 0,0)$, $(10 ~\rm m, 0, 2 ~\rm m)$, while the devices are uniformly and randomly distributed in a circle centered at $\left( {10~{\rm{m}},0,0 } \right)$ with a radius $3~\rm m$. The distance-dependent path loss model is given by $L\left( d \right) = {c_0}{\left( {d/{d_0}} \right)^{ - \alpha }}$,
where ${c_0} = -30 ~{\rm dB}$ is the path loss at the reference distance $d_0=1$ m,  $d$ is the link distance, and $\alpha$ is the path loss exponent. 
 We assume that  the HAP-IRS link and  the IRS-device link    follow Rician fading with the Rician factor of $3~\rm dB$ and the  path loss exponent of $2.2$, while the  HAP-device link follows Rayleigh fading   with the path loss exponent of $3.8$. We model the SI channel  ${\bf H}_{\rm SI}$   following  Rician probability distribution \cite{nguyen2014on}, i.e., ${\cal CN}\left( {\sqrt {\gamma \bar K/\left( {1{\rm{ + }}\bar K} \right)} {{{\bf{\bar H}}}_{{\rm{SI}}}},\gamma /\left( {1{\rm{ + }}\bar K} \right){{\bf{I}}_{{N_r}}} \otimes {{\bf{I}}_{{N_t}}}} \right)$, where ${\bar K}$ denotes its Rician factor and ${{{{\bf{\bar H}}}_{{\rm{SI}}}}}$   denotes  a deterministic matrix (we model all the elements in ${{{{\bf{\bar H}}}_{{\rm{SI}}}}}$ to be one in simulations), and $\gamma$  is used  to parameterize 
 the capability of  SI cancellation (SIC).
 We assume that all the devices have the same configurations with the non-linear EH model parameters  given by  $a_k=150$, $b_k=0.014$, and $M_k=0.024, \forall k$ \cite{lu2019global}.  In addition,  the weighting factors are assumed to be the same with ${\omega _k}=1$ for $k\in {\cal K}_2$, and thus we use the term  ``sum throughput'' instead of the term  ``weighted sum throughput'' in the simulations.
 Other system parameters are set as follows:  $N_t=N_r=4$, $N_d=2$, ${\bar K}=1$, $T=1~s$, $\sigma^2=-90~{\rm dBm}$,  and  $\varepsilon=10^{-2}$. 
 Furthermore, as  unveiled  in \cite{hua2021joint},  the device scheduling order has a negligible impact on the  FD-WPCN. Thus,  we adopt the increasing order
 scheduling where the device with the highest SNR of the HAP-device link is scheduled to
 transmit first as our scheduling strategy in the following simulations.

  \begin{figure}[!t]
	\centerline{\includegraphics[width=3.5in]{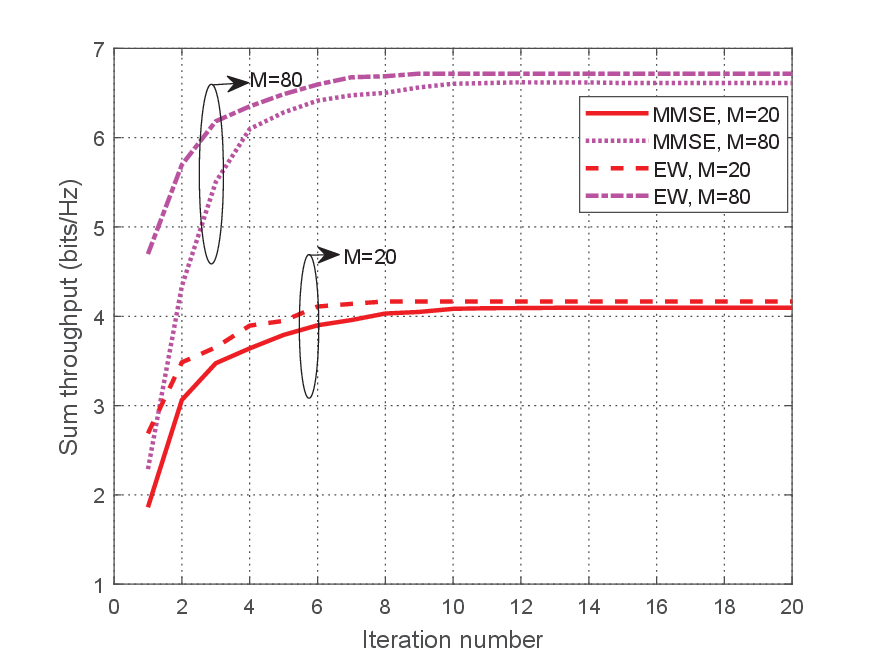}}
	\caption{ Convergence behaviors of  the proposed MMSE-based algorithm and EW-based algorithm.} \label{convergence} 
	\vspace{-0.5cm}
\end{figure} 
%\subsubsection{Comparison between EW-based Algorithm and MMSE-based Algorithm} 
Before the performance comparison, we first show the convergence behaviors of the proposed two  algorithms as shown in Fig~\ref{convergence}. In particular, we show the sum throughout  versus the number of iterations for the different number of IRS reflecting
elements, namely $M=20$ and $M=80$, with  $\gamma=-120 ~{\rm dB}$,  $P_{\max}=40~{\rm dBm}$, and $K=5$. It is observed that with the same $M$, both two algorithms nearly have the same number of iterations required for reaching convergence, while in each iteration, the complexity of the EW-based algorithm is much larger than that of the MMSE-based algorithm analyzed   in Section III-D and Section IV-F. In addition, we observe that the  EW-based algorithm 
achieves higher sum throughout than the MMSE-based algorithm. This can be explained  as follows. On the one hand,  problem  \eqref{p3} formulated  based on the MMSE method is equivalent to the original 
problem  \eqref{p1} only when the optimal solution is achieved, while the suboptimal  solution   of  problem  \eqref{p3} is not  guaranteed to be   the suboptimal   solution for   problem  \eqref{p1}, which inevitably  incurs a certain   performance loss.
On the other hand,   the MMSE-based algorithm has  additional two  blocks to be optimized compared to the EW-based algorithm, which indicates that  it is prone to getting trapped at undesired suboptimal solution due to the more stringent coupling among   variables.
 Furthermore, we observe that the performance gap between the two algorithms is small, even when $M$  becomes large. Therefore,
if not otherwise specified, the mentioned  schemes  in the following  are solved by the MMSE-based algorithm.

In order to show the performance gain brought by the IRS in the MIMO FD-WPCN,  we compare the following   schemes. 1) EW, no SI:  Our proposed scheme, which is solved by Algorithm~\ref{alg1} with perfect  SIC, i.e.,  $\gamma=0$; 2) MMSE, no SI ($\gamma=-120 ~{\rm dB}$): Our proposed scheme, which is solved by Algorithm~\ref{alg2} with $\gamma=0$ ($\gamma=-120 ~{\rm dB}$);
% 3) $\gamma=-120 ~{\rm dB}$/$\gamma=-90 ~{\rm dB}$: We consider two imperfect SIC  cases, namely $\gamma=-120 ~{\rm dB}$ and $\gamma=-90 ~{\rm dB}$, which are solved by Algorithm~\ref{alg2}; 
 3) PDBF 1:  We consider a \textit{partially dynamic IRS beamforming} (PDBF) scheme, where one common  IRS phase-shift vector is applied for DL WET, while the  phase-shift vectors vary with each time  slot for UL WIT.
4) Fixed time:  The time allocation for both DL WET
and UL WIT is  equally divided  into  $2K$ time slots with each of duration given by    $T/2K$; 5) Random IRS:
Each phase shift at the IRS is random and follows the
uniform distribution over $\left[ {0,2\pi } \right)$; 6) Without IRS: Without adopting the IRS. 7) Linear: We
solve a problem similar to problem \eqref{p1} but with the linear EH model. Then, we apply  the obtained solution to the actual system
with the  non-linear EH model and check if   energy-causality  constraint \eqref{p1_const_1} is satisfied. If it  is not satisfied, we solve the resulting problem by optimizing   time allocation to make    constraint  \eqref{p1_const_1} satisfied. 8) HD: The HAP operates in the HD mode.

  \begin{figure}[!t]
	\centerline{\includegraphics[width=3.5in]{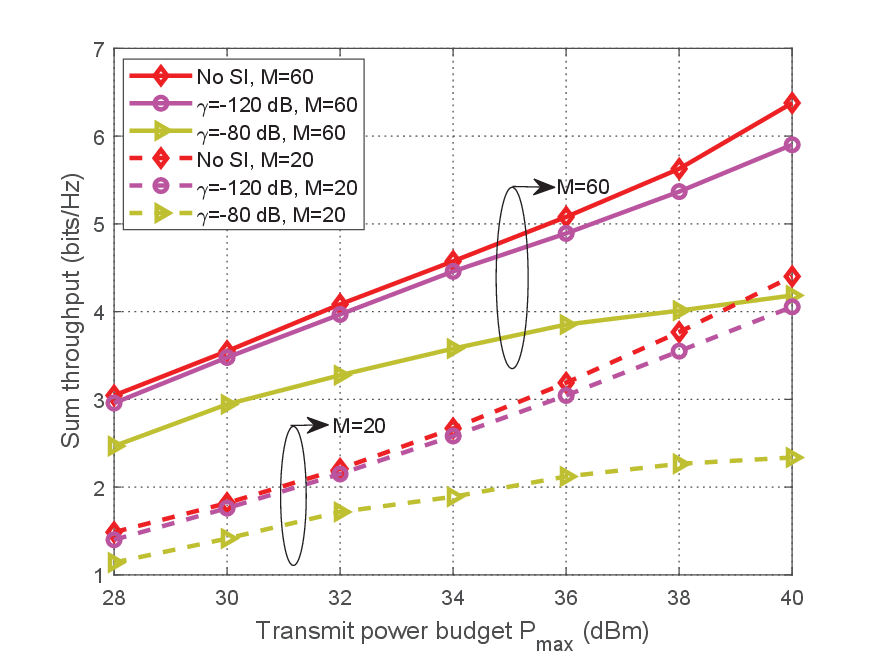}}
	\caption{Sum throughput versus    $P_{\max}$.} \label{vsgamma} 
	\vspace{-0.5cm}
\end{figure} 
\subsubsection{Effect of imperfect SIC }  
In  Fig.~\ref{vsgamma}, we study the  sum throughput versus $P_{\max}$ for different values of $\gamma$ with  $K=5$.
It is observed that when  $M=20$, the proposed scheme with $\gamma=-120 ~{\rm dB}$  achieves   nearly   the same  sum throughput as that with perfect SIC when $P_{\max}\le 34 ~{\rm dBm}$, even  when $P_{\max}$ becomes large, i.e., $P_{\max}\ge 34~ {\rm dBm}$, the performance gap between two schemes is still small. In addition, we  observe that  as $\gamma$ becomes larger,   the sum throughput will be smaller. This is expected since a higher value of  $\gamma$   incurs    stronger SI on the HAP receiver side,  which will degrade the system performance. Especially, when $\gamma=-80 ~{\rm dB}$ and $M=20$,  the sum throughput   increases slowly with    $P_{\max}$, even remains unchanged when $P_{\max}\ge 38~ {\rm dBm}$,  which indicates the importance of canceling the SI at the HAP. Similar results are also observed  for $M=80$.

 \begin{figure}[!t]
	\centerline{\includegraphics[width=3.5in]{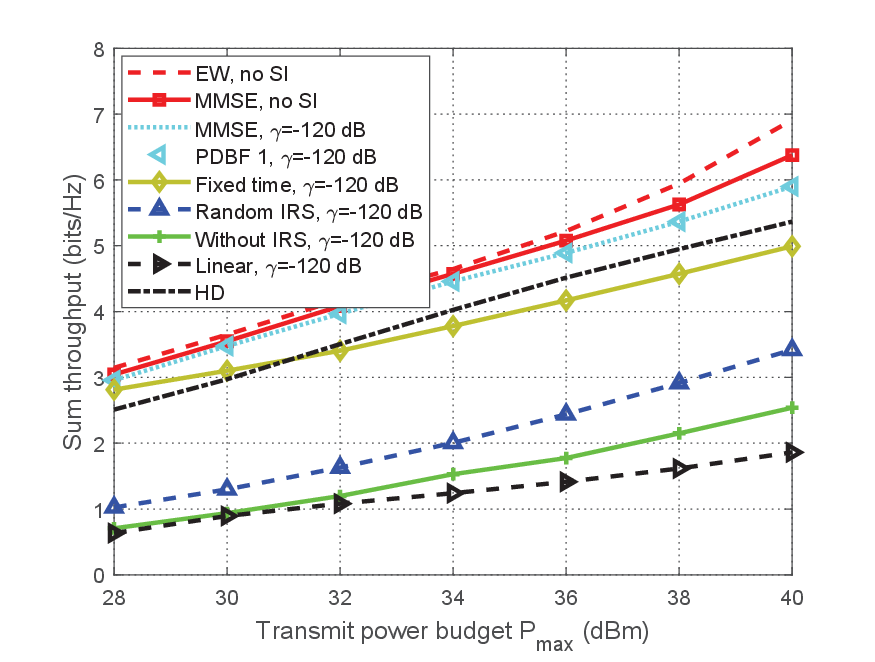}}
	\caption{Sum throughput versus   ${P}_{\rm max}$.} \label{vsP} 
	\vspace{-0.5cm}
\end{figure} 
\subsubsection{Sum Throughput Versus   Transmit Power} 
In Fig.~\ref{vsP}, we study the sum throughput obtained by all schemes versus  $P_{\max}$  with  $M=60$   and  $K=5$. It can be observed that the EW-based approach   achieves better performance than the  MMSE-based approach in terms of sum throughput, which is consistent with the  previous discussion.  It is also observed that the sum  throughput obtained by all  schemes monotonically increases with $P_{\max}$, even with the imperfect  SIC.
This can be understood  as follows. On the one hand, in the DL WET stage, i.e., $\tau_0$,  the  harvested energy by devices will significantly increase as   $P_{\max}$ increases, which  allows devices to transmit higher power   for UL WIT. On the other hand, in the UL WIT stage, i.e., $\tau_i, i\ne 0$,  we can  properly  decrease the HAP transmit power to alleviate the SI at the HAP receiver side  at the cost of transmitting less energy  by devices, which is also able to provide some certain performance gains. In addition, we can observe that our proposed scheme   outperforms the benchmark schemes with    fixed time, with  random IRS, and without IRS,  especially as  $P_{\max}$ increases, which demonstrates the benefits of the joint design of the IRS beamforming and resource allocation for improving the  FD-WPCN system performance. As expected, we observe that our proposed  FD-WPCN   significantly outperforms the HD-WPCN, especially when the SI is perfectly cancelled. Moreover, it is observed that the
scheme based on the linear EH model achieves less sum throughput 
compared to  schemes based on the non-linear EH
model due to the resource allocation mismatch.  Furthermore, one can observe that the ``PDBF 1'' scheme achieves the same performance with our proposed   scheme for the same $\gamma$, which indicates that the DBF for the dedicated DL WIT is not needed in practice (For $K=5$, the duration of DL WET is not zero, which has been   discussed in  Fig.~\ref{DLtime}). This is  because  although the harvested energy profiles  for devices via   dynamically adjusting the  IRS phase-shift vector across each sub-slot is different from that via only optimizing one common IRS phase-shift vector over the entire DL WET duration, the impact incurred by this difference   can be mitigated by optimizing the UL time allocation.

 \begin{figure}[!t]
	\centerline{\includegraphics[width=3.5in]{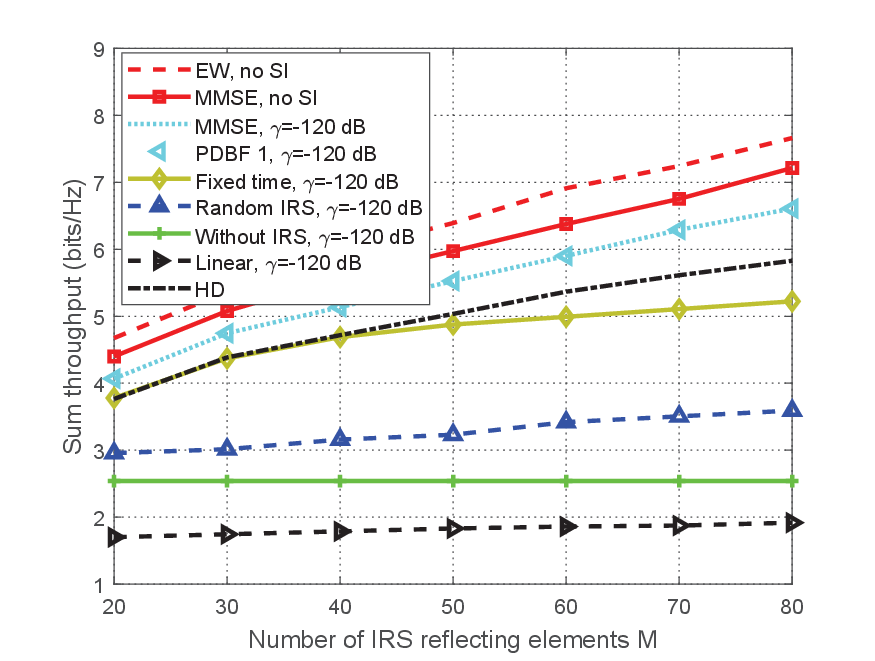}}
	\caption{Sum throughput versus   $M$.} \label{vsM}
	\vspace{-0.5cm} 
\end{figure} 
 \subsubsection{Sum Throughput Versus   Number of Reflecting Elements} 
 In  Fig.~\ref{vsM}, we compare the sum throughput obtained by all schemes versus   $M$ with  $P_{\max}=40~{\rm dBm}$  and  $K=5$.  It is first observed that the sum throughput obtained by  schemes with IRS phase shift optimization monotonically increases with $M$ since more reflecting elements installed at the IRS help achieve higher passive beamforming gain for both DL WET and UL WIT. 
 We observe that the EW-based algorithm with perfect SIC outperforms the other schemes, which again demonstrates the superiority of the EW-based algorithm  over the MMSE-based algorithm in terms of the sum  throughput. In addition, we observe  that with the same SI $\gamma=-120 ~{\rm dB}$, our proposed scheme achieves higher sum throughput than those schemes with   fixed time, with   random IRS, and without IRS, which  demonstrates the
 benefits of the  joint optimization of resource allocation and IRS phase shifts. Moreover,  the proposed FD scheme outperforms the    HD scheme in terms of sum throughput when the SI is well suppressed, i.e., $\gamma=-120 ~{\rm dB}$ or $\gamma=0$. Furthermore,  the performance gap between two schemes  would  be more pronounced when $M$ becomes large. This is because installing  more reflecting elements at the IRS will provide high passive beamforming gains and  more energy will be harvested by  devices under the FD mode than under  the HD mode, which allows devices to transmit higher power for  UL WIT.
 
% This is because as the HAP operates in an FD mode,   more energy will be harvested by devices and a higher transmit  power at devices   can be  used for improving the UL WIT rate. 
% 
% Interestingly, we observe that in the case where   the SIC is not well suppressed, i.e., $\gamma=-90 ~{\rm dB}$, the proposed FD scheme still  outperforms the HD scheme if  more reflecting elements at the IRS are adopted. This is because installing  more reflecting elements at the IRS will provide high passive gains so that 
% more energy will be harvested by the devices under the FD mode than  the HD mode, which thus achieves a higher sum throughput.

 \begin{figure}[!t]
	\centerline{\includegraphics[width=3.5in]{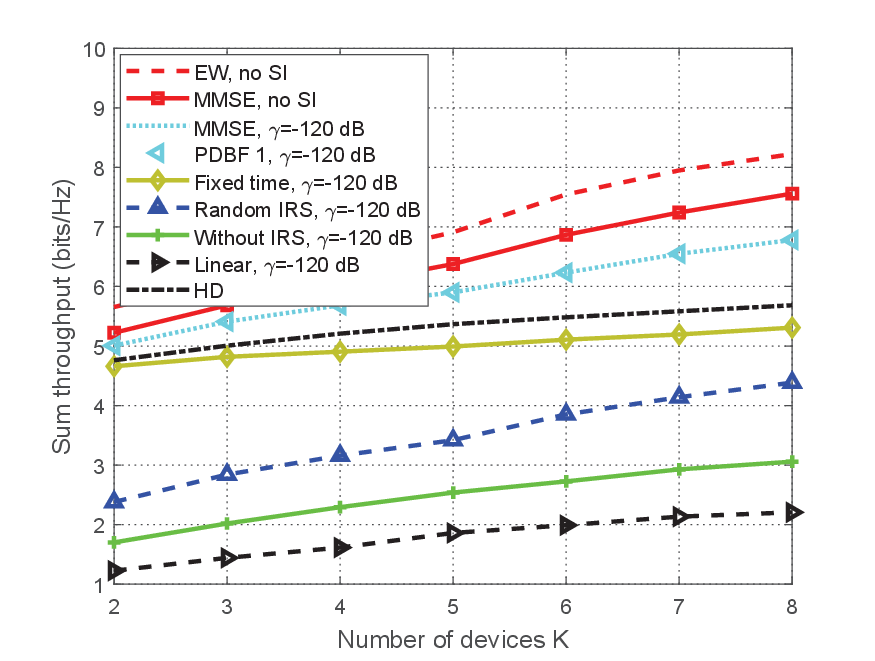}}
	\caption{Sum throughput versus   $K$.} \label{vsK} 
	\vspace{-0.5cm}
\end{figure} 
\subsubsection{Sum Throughput Versus   Number of Devices} 
In  Fig.~\ref{vsK}, we compare the sum throughput obtained by all schemes versus   $K$ with  $P_{\max}=40~{\rm dBm}$  and  $M=60$. It can be seen that the sum throughput obtained by all  schemes increases as $K$ increases due to the more multiuser diversity and higher design flexibility of IRS.
 In addition, as $K$ increases, the performance gap between our proposed scheme and the fixed time scheme becomes larger. This can be understood as follows. Since some of devices have poor channel conditions, it is not wise to allocate the time to these devices. In contrast, with the  optimization of  time allocation, we may not   assign any time to  devices with poor channel conditions for   UL WIT, while we  assign more  time to  devices with  good channel conditions for UL WIT. Therefore, the optimization of time allocation  plays an important role in system design, especially when $K$ is large. Moreover, we  can observe that our proposed scheme outperforms the other benchmark schemes, which again demonstrates the benefits of exploiting both IRS and FD in the MIMO WPCN. Furthermore, we can also observe  that with $\gamma=-120 ~{\rm dB}$, the ``PDBF 1'' scheme   achieves the same performance with our proposed scheme, which further demonstrates the non-necessity of adopting DBF for DL WET.
 \begin{figure}[!t]
	\centerline{\includegraphics[width=3.5in]{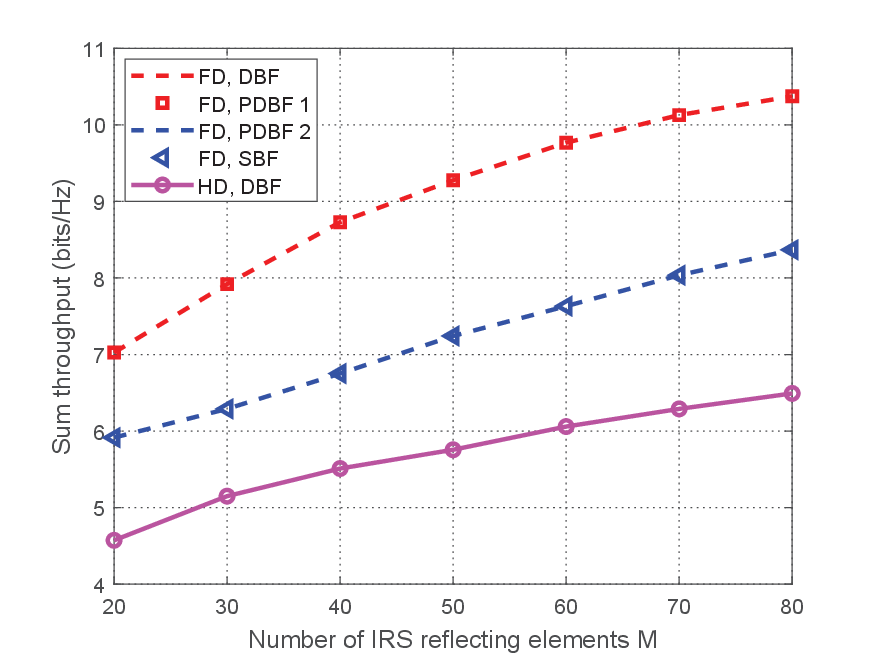}}
	\caption{Sum throughput versus   $M$ for three types of IRS
		configurations with $P_{\max}=40~{\rm dBm}$  and  $K=15$.} \label{dynamcivsstatic} 
\end{figure} 
 \begin{figure}[!t]
	\centerline{\includegraphics[width=3.5in]{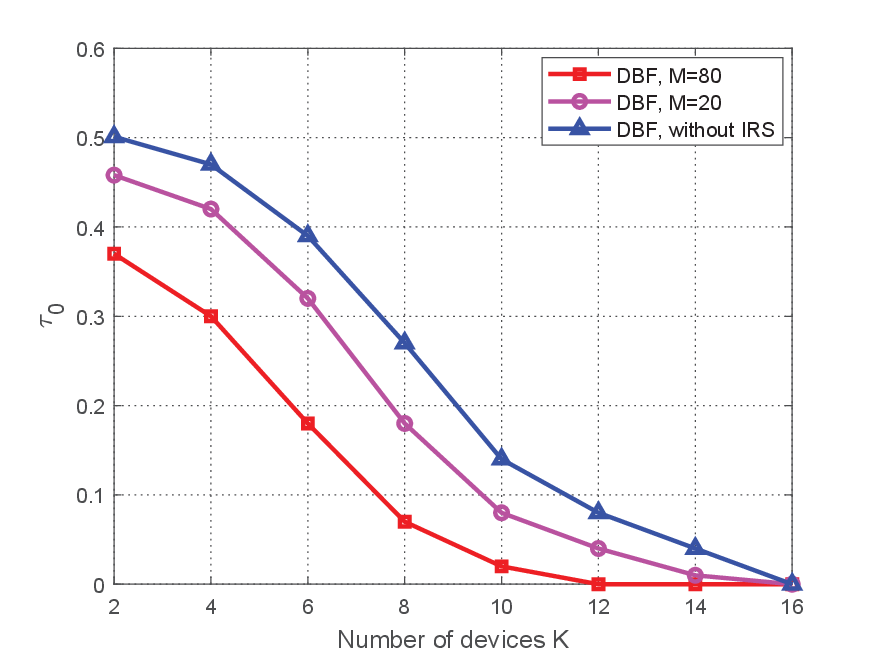}}
	\caption{Duration of DL WET  versus   $K$  with $P_{\max}=40~{\rm dBm}$.} \label{DLtime} 
	\vspace{-0.5cm}
\end{figure} 
 \subsubsection{Dynamic versus Static IRS Beamforming} 
  In  Fig.~\ref{dynamcivsstatic}, we study three types of IRS beamforming configurations in terms of
 sum throughout to strike a balance between the system performance and signaling overhead as
 well as implementation complexity. We compare the following schemes. 1) FD, DBF: Our proposed scheme; 2) FD, PDBF: There are two special cases for the PDBF scheme. For the first PDBF case, i.e.,  PDBF 1, which has been previously discussed. For the second PDBF case, i.e.,  PDBF 2, where there are two different IRS phase-shift vectors during the whole transmission  period with one for DL WET and the other for UL WIT; 3) FD, SBF: This is the \textit{static IRS beamforming} (SBF) scheme, where one common IRS  phase-shift vector is applied for both DL WET and UL WIT. 4) HD, DBF: Compared to the ``FD, DBF'' scheme, the HAP operates in the HD mode. First,  it is observed that the ``FD, DBF'' scheme achieves the same system performance with the ``FD, PDBF 1'' scheme. This is because as $K$ becomes large, i.e., $K=15$, the duration of DL WET time slot, i.e., $\tau_0$,  becomes almost zero. To under it more clearly, we study the    duration of DL WET time slot $\tau_0$ versus $K$ as shown in Fig.~\ref{DLtime}. 
  It is observed  that $\tau_0$ monotonically decreases  as   $K$ increases.  This is because when   $K$ is small, the number of time slots for EH during UL WIT is also small and thus the dedicated DL WET time slot plays an important role in improving system performance, 
  while   as $K$ becomes large,  the number of time slots for UL WIT becomes large, which indicates that the devices can still harvest sufficient energy even  without    the   dedicated DL WET time slot.   Similar results can also be  observed by comparing the  ``FD, PDBF 2'' scheme  with the ``FD, SBF'' scheme. In addition, we can observe that the ``FD, DBF'' scheme (or ``FD, PDBF 1'' scheme)  significantly outperforms the ``FD, PDBF 2'' scheme (or ``FD, SBF'' scheme). This is expected since for the former, the IRS is able to proactively
 generate artificial time-varying channels over each time slot to adapt to UL WIT so as to
 improve the system performance.

\section{Conclusion}
In this paper, we studied the joint design of passive beamforming   and resource allocation for the IRS-aided MIMO FD-WPCN by taking  into    account the  finite SI and the non-linear EH model. Specifically, the DL/UL time allocation,  precoding matrices,  transmit covariance matrices, and   phase shifts  were jointly optimized to maximize the sum throughput of the IRS-aided MIMO FD-WPCN.  We proposed two algorithms, namely EW-based algorithm and MMSE-based algorithm,  to achieve a balance between the system performance and the
computational complexity.  Simulation results  demonstrated the benefits of the    IRS  used for enhancing the performance of the MIMO FD-WPCN, especially when the  DBF was adopted. Besides, the  results revealed that the  FD mode is more beneficial
than the HD mode  when the SI is effectively suppressed.   Furthermore, the  results also showed that the  DBF for DL WET  may not be necessary  in practice.	
This work can be further extended by considering imperfect CSI, multiple IRSs, and quality-of-service  constraints  at devices, etc., which will be left as future work.

\bibliographystyle{IEEEtran}
\bibliography{FD_MIMO}
\end{document}